\documentclass[prd,twocolumn,preprintnumbers]{revtex4-1}

\usepackage{amsmath}
\usepackage{epsfig}
\usepackage{graphicx}
\usepackage{color}
\usepackage[normalem]{ulem}
\usepackage{url}
\usepackage{dirtytalk}
\usepackage{float}
\usepackage[breaklinks, plainpages=false, colorlinks=true, anchorcolor=cyan, linkcolor=red, citecolor=cyan, urlcolor=magenta, bookmarks=false]{hyperref}

\usepackage[caption=false]{subfig}

\begin{document}
\renewcommand{\thefigure}{\arabic{figure}}
\setcounter{figure}{0}

 \def\I{{\rm i}}
 \def\E{{\rm e}}
 \def\D{{\rm d}}

\bibliographystyle{apsrev}

\title{Rapid inference for individual binaries and a stochastic background with pulsar timing array data}

\author{Aiden Gundersen}
\affiliation{eXtreme Gravity Institute, Department of Physics, Montana State University, Bozeman, Montana 59717, USA}

\author{Neil J. Cornish}
\affiliation{eXtreme Gravity Institute, Department of Physics, Montana State University, Bozeman, Montana 59717, USA}

\begin{abstract} 
The analysis of pulsar timing array data has provided evidence for a gravitational wave background in the nanohertz band. This raises the question of what is the source of the signal, is it astrophysical or cosmological in origin? If the signal originates from a population of supermassive black hole binaries, as is generally assumed, we can expect to see evidence for both anisotropy and to be able to resolve signals from individual binaries as more data are collected. The anisotropy and resolvable systems are caused by a small number of loud signals that stand out from the crowd. Here we focus on the joint detection of individual signals and a stochastic background. While methods have previously been developed to perform such an analysis, they are currently held back by the cost of computing the joint likelihood function. Each individual source is described by $N=8+2N_p$ parameters, where $N_p$ are the number of pulsars in the array. With the latest combined datasets having over one hundred pulsars, the parameter space is very large, and consequently, it takes a large number of likelihood evaluations to explore these models. Here we present a new approach that extends the Fourier basis method, previously introduced to accelerate analyses for stochastic signals, to also include deterministic signals. Key elements of the method are that the likelihood evaluations are per-pulsar, avoiding expensive operations on large matrices, and the templates for individual binaries can be computed analytically or using fast Fourier methods on a sparsely sampled grid of time samples. The net result is an analysis that scales better than quadratically with the size of the dataset, while the approach currently being used in most analyses scales quartically or worse with the number of data points. As datasets grow with more observations, this analysis will be orders of magnitude faster than previous approaches.
\end{abstract}

\maketitle

\section{Introduction}\label{intro}

Pulsar timing arrays (PTAs) measure the time-of-arrival (TOAs) of radio pulses produced by millisecond pulsars. By recording these TOAs for decades, PTAs are sensitive to gravitational waves (GWs) with nHz frequencies. At the time of writing this paper, evidence for a stochastic gravitational wave background (GWB) has been found by the North American Nanohertz Observatory for Gravitational Waves (NANOGrav), the Parkes Pulsar Timing Array (PPTA), the European Pulsar Timing Array (EPTA), and the Chinese Pulsar Timing Array (CPTA)~\cite{NG15year, EPTAGWB, ParkesPTA, ChinesePTA}. With this evidence, the next question is: through what process does the GWB signal originate? The most common belief, consistent with observed data, is the background is realized by a population of supermassive black hole binaries (SMBHBs). These binaries form as a result of galaxy mergers and emit GWs in the nHz band~\cite{NG15year}. However, many cosmological origins for the signal have also been proposed~\cite{NANOGrav:2023hvm,EPTA:2023xxk}. If the GWB is realized through a population of SMBHBs, then anisotropy in the background and individual binaries should become resolvable as more TOAs are observed~\cite{Rosado:2013wva,Becsy:2022pnr}.

The observed TOAs are not determined solely by GW signals. Deterministic delays such as pulsar spin period, spin derivative, and proper motion significantly contribute to the radio pulse TOAs. A timing model has been constructed per pulsar that predicts the TOAs to within $\mathcal{O} (1\,\mu\text{s})$ accounting for such deterministic contributions~\cite{NGTiming}. Millisecond pulsars additionally exhibit red noise (RN) due to quasi-random-walks in their pulse phase, period, and spindown rate arising from intrinsic instabilities~\cite{red_noise}. Moreover, dispersion of the radio pulses in the interstellar medium induces frequency-dependent time delays. The variation of this dispersion measure (DM) sources additional red noise in the TOA observations. White noise from instrumental effects, radiometer noise, and pulse phase jitter also populate the TOAs~\cite{NGnoisebudget}. Each of these noise and signal processes are, to some degree, covariant with one another and with GW signals. Therefore, a joint PTA analysis must model all these signal and noise processes simultaneously.

Pulsar timing data is unevenly sampled and the noise is heteroscedastic, necessitating the analysis to be carried out in the time-domain. As more TOAs are recorded the analysis becomes increasingly computationally expensive, by virtue of an expensive likelihood function which must be evaluated many times over the parameter space. The problem is worsened when one considers high-dimensional models with large parameter volume, increasing the number of likelihood evaluations required.

The likelihood evaluation can be made significantly more efficient by representing the GWB and intrinsic pulsar RN in a Fourier basis, as first presented by Lentati et. al.~\cite{Lentati:2012xb}. The Fourier coefficients which describe the signal and noise then become model parameters and must be sampled over, greatly increasing the dimension of the model. The likelihood in this form however is hyper-efficient which could speed up the analysis if the high-dimensional parameter space is explored effectively. 

A general result, which is not widely appreciated, is that for stochastic signals ``subtraction equals division''~\cite{Lentati:2012xb, Cornish:2013nma,Romano:2016dpx}. That is, we can either subtract stochastic signals and noise from the data, or instead account for them in the inverse covariance matrix which appears in the likelihood. For stationary stochastic processes it is natural to use a frequency-domain description, and the subtraction can be performed using a basis of sines and cosines. For Gaussian stochastic processes, the amplitudes of the sines and cosines follow a multivariate normal distribution. The amplitudes can be analytically marginalized over (integrated out), resulting in a modified covariance matrix in the marginalized likelihood. This effectively replaces the signal and noise subtraction by division. 

In usual PTA analyses, the Fourier coefficients describing the GWB and RN are analytically marginalized over. This results in a dense covariance matrix which must be inverted for every likelihood evaluation. By including the Fourier coefficients as model parameters and fixing the white noise model, the covariance matrix is constant and its inverse along with relevant inner products can be stored for use in every future likelihood evaluation. There are no expensive matrix inversions in the likelihood when modeling the Fourier coefficients. The drawback to this approach is a high-dimensional model with $\mathcal{O}(10^3)$ model parameters for realistic datasets. It's therefore crucial that efficient sampling techniques be employed to explore the large parameter volume.

Recently, there has been renewed interest in the Fourier basis approach, with several studies looking at performing what amounts to a ``Bayesian Fourier transform'', to first produce posterior distributions for the Fourier coefficients, which can then be used to model the signals and noise in a hierarchical Bayesian analysis~\cite{Laal:2023etp,Laal:2024hdc,RF}. Our approach is a little different. The inter-pulsar correlations expected of a stochastic GWB are always included, but we may simultaneously impose a spectral model or not. If we do not model the spectrum initially, we can use our inference on the Fourier coefficients to analyze the spectrum in a subsequent second stage analysis. In other words, the spectra of signal and noise processes can be modeled jointly with the Fourier coefficients via a hierarchical Bayesian scheme, or we can break the analysis into two stages and use the initial recovery of the Fourier coefficients as data for the second stage.

In this paper, we extend the approach from Lentati el. al.~\cite{Lentati:2012xb} to include deterministic signals. Generally, the Fourier coefficients can be separated corresponding to the signals they represent: the stochastic GWB, the intrinsic pulsars RN, and deterministic signals. As the GWB and RN are stochastic processes, we include their respective Fourier coefficients as model parameters and sample over them in the analysis (as opposed to analytically marginalizing them out of the model). The TOA-delays induced by deterministic signals however can be evaluated analytically over any choice of time samples, given some set of deterministic model parameters. Then we can represent the deterministic signal with a Fourier basis by either performing an analytic Fourier transform or a discrete fast Fourier transform (FFT). That is, we do not sample over the Fourier coefficients which represent the deterministic signal. Instead, we sample over the usual deterministic model parameters, and only ``under the hood'' use its representative Fourier coefficients obtained through a Fourier transform to retain the hyper-efficient likelihood evaluation.

The cost of exploring a model using say, an efficient Markov Chain Monte Carlo (MCMC) algorithm, scales somehwere between lineary and quadratically with the number of parameters. For example, the exploration of posteriors that follow a multivariate normal distribution scales quadratically using a naive random-walk Metropolis~\cite{10.1214/aoap/1034625254}, linearly using ideal Gibbs sampling, and as the five-fourths power using Hamiltonian sampling~\cite{10.3150/12-BEJ414,Freedman:2022tnf}. The number of parameters in our GWB and RN model scales linearly with the product of the number of pulsars and the number of Fourier coefficients. This number grows as more pulsars are added to the array and as the time span of the dataset increases, necessitating more terms in the Fourier expansion. In contrast, the size of the covariance matrices that appear in the likelihood when the Fourier coefficients are integrated out grow quadratically with the number of data points, and the cost of inverting these matrices cubically with the number of data points~\cite{NANOGrav:2023icp}. Then there is the additional cost of sampling the posterior with this expensive likelihood function.
The number of data points grows as the product of the number of pulsars and the duration of the observations. In short, our approach scales better than quadratically with the size of the dataset, while the approach currently being used in most PTA analyses scales quartically or worse with the number of data points.

\section{Continuous wave searches}

Gravitational waves originating from individual SMBHBs are approximately monochromatic, and for this reason often called continuous wave (CW) signals, see Section~\ref{ind_SMBHB} for a discussion. Joint PTA analyses search for one or more CW signals while simultaneously modeling the stochastic GWB, other deterministic signals, and noise processes. Some methods make the simplifying assumption that correlations induced by a stochastic GWB can be ignored~\cite{Babak:2011mr,Ellis:2013hna,QuickCW}. Other methods have been developed that include inter-pulsar correlations~\cite{Becsy:2019dim,NGCWsearch,EPTACWsearch}, but these approaches are currently held back by the computational cost. In some analyses the joint model is simplified to make the runtime feasible. For example, if an individual binary has an electromagnetic counterpart, then the parameter space for the CW signal can be constrained in some of its dimensions, say sky location~\cite{targetedsearch}. Presently, NANOGrav and the EPTA have found weak evidence for CWs using various joint analysis techniques~\cite{NGCWsearch,EPTACWsearch}.

The NANOGrav search for CWs uses a resampling procedure~\cite{NGCWsearch}. The stochastic GWB was first modeled as a common uncorrelated red noise (CURN) process, and the inter-pulsar correlations expected from the GWB were applied in post-processing. Importance weights were calculated using the ratio of likelihoods with and without the inter-pulsar correlations on a thinned set of posterior samples. Sampling according to these weights can produce the posterior for a joint model including inter-pulsar correlations in the GWB. However relatively few samples survive this process, so the results are not fully robust. The EPTA has also found weak evidence for a CW~\cite{EPTACWsearch}, sampling the likelihood constructed by \texttt{ENTERPRISE}. This likelihood relies on the inversion of dense covariance matrices, and the analysis will not be tractable as the number of observations increase.

\texttt{QuickCW} is a fast analysis for CW + CURN models~\cite{QuickCW}, commonly used in joint analyses. It precomputes and stores filters used in the likelihood for a given set of parameters, and these filters need only be recomputed when a subset of parameters are updated. This leads to a blocked sampling scheme which can efficiently conduct a joint analysis. However, \texttt{QuickCW} does not model inter-pulsar correlations we expect from a stochastic GWB. If such correlations were included the filters would have to be recomputed every time the background model parameters were updated and the filters in each pulsar would be correlated according to the background. Because the filters are correlated, inner products between filters are no longer diagonal in pulsar space and a dense $4N_p\times 4N_p$ correlation matrix must be computed frequently in the analysis, where $N_p$ is the number of pulsars in the array.

In this paper, we present methods which jointly model the stochastic GWB, RN intrinsic to pulsars, and deterministic signals such as CWs. This method includes the inter-pulsar correlations expected in the background through hierarchical modeling, so no post-processing is required. Moreover, the covariance matrix is constant when the white noise model is fixed, and its inverse along with relevant inner products can be precomputed and stored for use in a hyper-efficient likelihood evaluation.

\section{Signal Model and Likelihood}\label{model}

We generalize the signal model from Ref.~\cite{Lentati:2012xb} to include deterministic sources. The extension is fully general, and can be applied to any deterministic signal. The pulse arrival timing data, ${\bf t}$, are made up of contributions from red and white noise, ${\bf n}_R,{\bf n}_W$, deterministic timing delays ${\bf t}_T$, a stochastic gravitational wave background ${\bf t}_B$, and individual deterministic signals ${\bf t}_D$. The timing residuals are found by subtracting the reference timing model: $\delta {\bf t} = {\bf t} - {\bf t}_T$.  The reference timing model is constructed pulsar by pulsar, and does not account for the presence of the gravitational wave signals common to all the pulsars. Because the reference timing model can absorb some of the signals, it is necessary to adjust the timing model when performing a joint analysis of all the data. Under the assumption that the gravitational wave signals produce a small perturbation to the timing model, the correlations are accounted for by linearizing about the reference model:
\begin{equation}
\delta {\bf t} ' = \delta {\bf t}  - {\bf M} \boldsymbol{\epsilon}\, ,
\end{equation}
where $\boldsymbol{\epsilon}$ are the linear deviations to the timing model parameters and ${\bf M}$ is the timing design matrix~\cite{Gmatrix, Gmatrix1, Gmatrix2}.  Going one step further and subtracting the signal and red noise model we are left with the white noise timing residuals
\begin{equation}
{\bf r}  = \delta {\bf t}  -  {\bf M}\boldsymbol{\epsilon} -  {\bf t}_B - {\bf t}_D - {\bf n}_R\, .
\end{equation}

Assuming the white noise residuals are Gaussian distributed with zero mean and noise covariance matrix ${\bf N}$, the likelihood function is
\begin{equation}\label{wn_like}
p( {\bf r} | \boldsymbol{\lambda})  = \sqrt{{\rm det} \left( ( 2 \pi {\bf N})^{-1}\right)} \, e^{-\frac{1}{2} {\bf r}(\boldsymbol{\lambda})^T {\bf N}^{-1} {\bf r}(\boldsymbol{\lambda})} \, .
\end{equation}
Here $\boldsymbol{\lambda}$ denotes all the parameters in the timing model, signal model and red noise model. At this stage, the likelihood can be factored into the product of the likelihoods for each pulsar.  

Generalizing the treatment of Ref.~\cite{Lentati:2012xb}, we express the stochastic GWB, the deterministic signals, and the intrinsic pulsar RN in terms of a Fourier basis ${\bf F}$, with entries
\begin{equation}
F_{k t} = \left\{ \sin\left(\frac{2 \pi k}{T} t\right), \cos\left(\frac{2 \pi k}{T} t\right) \right\} \, .
\end{equation}
In standard analyses, $T$ is the time span of the longest observed pulsar in the array. However $T$ may be chosen longer than the observation span~\cite{lowrankapprox}. When modeling deterministic signals, we choose $T$ to be nearly twice the observation span, see Section~\ref{sec:DetSigFR} and Appendix~\ref{app:F_CW}. The sample times $t$ are different for each pulsar, and are unevenly spaced. The discrete frequencies of the Fourier basis are indexed by the integer $k$. The stochastic GWB, deterministic signal, and RN need not share a common Fourier basis; each process is allowed a unique period, $T$, and maximum frequency bin, $k_\text{max}$, to set the basis. The Fourier representation of each signal is related to its time-domain counterpart as $\mathbf{t}_B=\mathbf{F}_B\mathbf{a}_B$, $\mathbf{t}_D=\mathbf{F}_D\mathbf{a}_D$, and $\mathbf{n}_R=\mathbf{F}_R\mathbf{a}_R$. Under a shared basis, the timing residuals are
\begin{equation}
{\bf t}_B  + {\bf t}_D +  {\bf n}_R = {\bf F} {\bf a} 
\end{equation}
where the Fourier coefficients ${\bf a}$ have contributions from the three terms:
\begin{equation}
{\bf a} = {\bf a}_B  + {\bf a}_D +  {\bf a}_R \, .
\end{equation}

\subsection{Stochastic processes in a Fourier representation}

The stochastic Fourier coefficients are approximately described by zero mean, Gaussian distributions with covariance matrices
\begin{eqnarray} \label{cov}
C^B_{Ii,Jj} &=& {\rm E}[a^B_{Ii}a^B_{Jj}] = \alpha_{IJ} \rho_i \delta_{ij}  \nonumber \\
C^R_{Ii,Jj} &=& {\rm E}[a^R_{Ii}a^R_{Jj}] =  \delta_{IJ} \kappa_{Ii} \delta_{ij} \, ,
\end{eqnarray}
where ${I,J}$ labels the pulsar and ${i,j}$ refers to the discrete frequencies. $\rho_i$ denotes the power spectrum of the stochastic GWB, and $\kappa_{Ii}$ denotes the power spectrum of the RN in the $I^{\rm th}$ pulsar. There is no sum over the repeated indices. The pulsar RN is assumed to be uncorrelated between pulsars, while the GWB follows the Hellings-Downs correlation pattern~\cite{1983ApJ...265L..39H}
\begin{equation}
 \alpha_{IJ}  = \frac{3}{2} \beta_{IJ} \ln \beta_{IJ} -\frac{1}{4} \beta_{IJ} + \frac{1}{2} + \frac{1}{2} \delta_{IJ}
 \label{eq:HD-corr}
\end{equation}
where $\beta_{IJ} = (1-\cos\theta_{IJ})/2$, and $\theta_{IJ}$ is the angle between pulsars $I$ and $J$ on the sky.

In PTA datasets, the power spectrum for the intrinsic pulsar RN and stochastic GWB are usually described by a free spectral or power law model~\cite{NGnoisebudget}, although there are other parameterizations. The free spectral model allows each element of the power spectrum, $\rho_i$ or $\kappa_{Ii}$, to be a free parameter. The power law uses an amplitude $A$ and spectral index $\gamma$ to parameterize the power spectrum as
\begin{equation}\label{eq:powerlaw}
    \rho_i \,(A, \gamma) = \frac{A^2}{12\pi^2}\frac{1}{T}\bigg(\frac{f_i}{1\,\text{yr}^{-1}}\bigg)^{-\gamma}\,\text{yr}^2
\end{equation}
where $f_i = i/T$ and we choose a reference frequency $f_\text{ref} = 1\,\text{yr}^{-1}$. The power law for intrinsic RN is modeled similarly, but each pulsar is allowed its own amplitude and spectral index.

Rather than modeling two independent sets of Fourier coefficients for the GWB and RN, we may combine their respective coefficients if they share a Fourier basis, $\mathbf{a}_{R+B} = \mathbf{a}_R + \mathbf{a}_B$. That is, the background and intrinsic pulsar noise are considered one stochastic process modeled using one set of Fourier coefficients. Because the sum of two Gaussian processes is another Gaussian process, the combined coefficients are described by a zero mean, Gaussian distribution with covariance matrix
\begin{equation}
    C^{R + B}_{Ii,Jj} = E[a^{R+B}_{Ii}a^{R+B}_{Jj}] = \delta_{IJ}\kappa_{Ii}\delta_{ij} + \alpha_{IJ}\rho_i\delta_{ij}\,.
    \label{combined-cov}
    \end{equation}

Modeling DM variations as a chromatic red noise Gaussian process would require us to use a Fourier basis distinct from that of other red noise processes. Specifically, the Fourier design matrix for DM variations is scaled by an additional $1 / \nu_{\text{obs}, i}^2$ factor where $\nu_{\text{obs}, i}$ is the radio frequency of the $i^\text{th}$ observed TOA~\cite{NGnoisebudget}. DM variations are modeled by a distinct set of Fourier coefficients which must be independently sampled, increasing the dimensionality of the parameter space. More efficient sampling techniques for the coefficients, such as Gibbs sampling~\cite{Laal:2023etp, Laal:2024hdc}, may mitigate the difficulties of increased dimensionality. In this paper, we ignore red noise processes from DM variations. 

The stochastic models above assume the Fourier modes are orthogonal, which corresponds to assuming the stochastic GWB and the intrinsic pulsar RN are stationary and the data observation spans an infinite period of time. Because the data is unevenly sampled in the time-domain, inference of the mode amplitudes will show correlations between different frequencies~\cite{Lentati:2012xb,van2014new,Allen:2024uqs}. Moreover, a finite observation period is equivalent to applying a rectangular window to the data, the Fourier transform of which is a cardinal sine function. The resulting Fourier transform of the signal is the convolution of the original Fourier expansion with the cardinal sine function. This induces additional correlations between different Fourier modes~\cite{BeyondDiagonalApprox, FreqCorrelations}. Such inter-frequency correlations are neglected in this paper. It is possible to include these finite-time correlations in the second stage of our hierarchical analysis, and this approach will be explored in future work.

\subsection{Deterministic signals in a Fourier representation}\label{sec:DetSigFR}

It's worth noting we need not represent the deterministic signal $\mathbf{t}_D$ in a Fourier basis. If the deterministic signal is computationally efficient to evaluate in the time-domain, then the white noise timing residuals should be written as
\begin{equation}
    {\bf r}  = \delta {\bf t}  -  {\bf M}\boldsymbol{\epsilon} -  \mathbf{F}_B\mathbf{a}_B - {\bf t}_D - \mathbf{F}_R\mathbf{a}_R\, .
\end{equation}
and the likelihood function is retains the form of Eq.~(\ref{wn_like}). However, in this paper we will assume the deterministic signal is a CW from an individual SMBHB, which is discussed in Section~\ref{ind_SMBHB}. Such a signal is nearly monochromatic, and can therefore be represented with relatively few Fourier coefficients. Rather than evaluating the deterministic CW model over every observed TOA, we need only sparsely sample the model in the time-domain to understand its frequency content. For this reason, a Fourier basis is a compressed representation of the CW signal and more computationally efficient than the explicit time-domain model.

As opposed to our assumption of the stochastic contributions, the coefficients of the deterministic signal model will include correlations between different frequencies. The coefficients are given by
\begin{equation} \label{dft}
{\bf a}_D(\boldsymbol{\zeta})  = {\cal F}\{{\bf h}(\boldsymbol{\zeta})\},
\end{equation}
where ${\cal F}$ denotes a discrete Fourier transform and ${\bf h}(\boldsymbol{\zeta})$ is the deterministic time-domain signal model, described by parameters $\boldsymbol{\zeta}$.

The deterministic signal model $\mathbf{h}(\boldsymbol{\zeta})$ can be evaluated over any choice of time samples $\mathbf{t}_m$ before the Fourier transform. The extent and density these time samples will determine the Fourier basis $\mathbf{F}_D$. Moreover, the accuracy of the Fourier representation, i.e. how closely $\mathbf{F}_D\mathbf{a}_D$ aligns with $\mathbf{h}(\boldsymbol{\zeta})$ evaluated over the observed TOAs, depends whether $\mathbf{t}_m$ was chosen to sufficiently capture the frequency content of the deterministic signal. For example, a deterministic signal with significant power in high frequency bins requires dense $\mathbf{t}_m$.

Because a CW is well represented using only a few Fourier coefficients, the time samples $\mathbf{t}_m$ can be sparse relative to the observed TOAs. If $\mathbf{t}_m$ is evenly spaced and the number of elements is a power of 2, a fast Fourier transform (FFT) efficiently computes the associated Fourier coefficients, $\mathbf{a}_D$. The subsequent application of the Fourier design matrix to these coefficients, $\mathbf{F}_D\mathbf{a}_D$, yields the deterministic model on the observed TOAs more efficiently than simply evaluating the model over the TOAs from the beginning. Essentially, the Fourier representation of the deterministic signal is nothing more than an interpolation. We evaluate the deterministic model over a sparse set of evenly spaced time samples, then interpolate the model onto the observed TOAs by composing the Fourier design matrix and the FFT.

For further discussion on choosing $\mathbf{t}_m$, including accuracy tests on the reconstructed waveforms and computation timing, see Appendix~\ref{app:F_CW}. It is possible to compute the discrete Fourier transform analytically for CW signals. The resulting expressions involve trigonometric functions, and it may be more efficient to compute the transform numerically using a FFT, depending on the desired number of Fourier modes.

\subsection{Individual black hole binaries}\label{ind_SMBHB}

The treatment here follows Ref.~\cite{Corbin:2010kt}. Gravitational wave signals can be expressed in terms of the tensor
\begin{equation}
	h_{ab}(t, \boldsymbol{\zeta}) = e_{ab}^{+}(\hat{\Omega}) \; h_+(t, \boldsymbol{\zeta}) + e_{ab}^{\times}(\hat{\Omega}) \; h_\times(t, \boldsymbol{\zeta}) \,,
\end{equation}
where $\hat{\Omega}$ is a unit vector from the GW source at sky location $(\theta,\phi)$ to the Solar System barycenter (SSB), 
$h_{+,\times}$ are the polarization amplitudes, 
and $e_{ab}^{+,\times}$ are the polarization tensors. 
The polarization tensors can be written in the SSB frame as
\begin{eqnarray}
	e_{ab}^{+}(\hat{\Omega}) &=& \hat{m}_a \; \hat{m}_b - \hat{n}_a \; \hat{n}_b \,, \\
	e_{ab}^{\times}(\hat{\Omega}) &=& \hat{m}_a \; \hat{n}_b + \hat{n}_a \; \hat{m}_b \,,
\end{eqnarray}
where
\begin{eqnarray}
	\hat{\Omega} &=& -\sin\theta \cos\phi \; \hat{x} - \sin\theta \sin\phi \; \hat{y} - \cos\theta \; \hat{z} \,, \\
	\hat{m} &=& \sin\phi \; \hat{x} - \cos\phi \; \hat{y} \,, \\
	\hat{n} &=& -\cos\theta \cos\phi \; \hat{x} - \cos\theta \sin\phi \; \hat{y} + \sin\theta \; \hat{z} \,.
\end{eqnarray}
The response of a pulsar to the source is described by the antenna pattern functions $F^+$ and $F^\times$
\begin{eqnarray}
	F^+(\hat{\Omega}) &=& \frac{1}{2} \frac{(\hat{m} \cdot \hat{p})^2 - (\hat{n} \cdot \hat{p})^2}{1+\hat{\Omega} \cdot \hat{p}} \,, \\
	F^\times(\hat{\Omega}) &=& \frac{(\hat{m} \cdot \hat{p}) (\hat{n} \cdot \hat{p})}{1+\hat{\Omega} \cdot \hat{p}} \,,
\end{eqnarray}
where $\hat{p}$ is a unit vector pointing from the Earth to the pulsar. The effect of a GW on a pulsar's residuals can be written as
\begin{equation}\label{signal}
	s(t,  \boldsymbol{\zeta}) = F^+(\hat{\Omega}) \; \Delta s_+(t, \boldsymbol{\zeta}) + F^\times(\hat{\Omega}) \; \Delta s_\times(t,  \boldsymbol{\zeta}) \,,
\end{equation}
where $\Delta s_{+,\times}$ is the difference between the signal induced at the pulsar and at the Earth 
(the so-called ``pulsar term'' and ``Earth term''), 
\begin{equation}
	\Delta s_{+,\times}(t, \boldsymbol{\zeta}) = s_{+,\times}(t_p,  \boldsymbol{\zeta}) - s_{+,\times}(t,  \boldsymbol{\zeta}) \,,
\end{equation}
where $t$ is the time at which the GW passes the SSB and $t_p$ is the time at which it passes the pulsar. 
From geometry, we can relate $t$ and $t_p$ by
\begin{equation}\label{pulsartime}
	t_p = t - L (1 + \hat{\Omega} \cdot \hat{p}) \,,
\end{equation}
where $L$ is the distance to the pulsar.

For a circular binary, at zeroth post-Newtonian (0-PN) order, $s_{+,\times}$ is given by
\begin{eqnarray}
    s_+(t,  \boldsymbol{\zeta}) &=& \frac{{\cal M}^{5/3}}{d_L \, \omega(t)^{1/3}} \left[ \sin 2\Phi(t) \, \left(1+\cos^2\iota\right) \, \cos2\psi \right. \nonumber \\
			&& \left. + 2 \cos 2\Phi(t) \, \cos \iota \, \sin 2\psi \right] \,, \label{eq:signal1} \\
	s_\times(t,  \boldsymbol{\zeta}) &=& \frac{{\cal M}^{5/3}}{d_L \, \omega(t)^{1/3}} \left[ -\sin 2\Phi(t) \, \left(1+\cos^2\iota\right) \, \sin2\psi \right. \nonumber \\
			&& \left. + 2 \cos 2\Phi(t) \, \cos \iota \, \cos 2\psi \right] \,, \label{eq:signal2}
\end{eqnarray}
where $\iota$ is the inclination angle of the SMBHB, $\psi$ is the GW polarization angle, 
$d_L$ is the luminosity distance to the source, 
and ${\cal M} \equiv (m_1 m_2)^{3/5}/(m_1+m_2)^{1/5}$ 
is a combination of the black hole masses $m_1$ and $m_2$ 
called the ``chirp mass.''  The frequency and phase evolution have the form
\begin{eqnarray}
    \omega(t) &=&  \left(\omega_0^{-\frac{8}{3}} - \frac{256}{5} {\cal M}^{\frac{5}{3}} t \right)^{-\frac{3}{8}} \nonumber \\	
	\Phi(t)  &=& \Phi_0 + \frac{1}{32 {\cal M}^{5/3}}  \left( \omega_0^{-\frac{5}{3}} - \omega^{-\frac{5}{3}} \right)\,.
\end{eqnarray}
To a good approximation, over the observation time $T$,  the frequency of the Earth term and pulsar term can be taken as constant, however the two frequencies can differ due to the projected time delay:
\begin{eqnarray}
    \omega_E &=& \omega(t_0) \nonumber \\	
    \omega_I &=& \omega(t_0- L_I (1 + \hat{\Omega} \cdot \hat{p}_I)) \nonumber \\
     & \approx &   \omega_E -  \frac{96}{5} \omega_E^{11/3} {\cal M}^{5/3} L_I (1 + \hat{\Omega} \cdot \hat{p}_I) \, .
\end{eqnarray}
This in turn means that the Earth terms and Pulsar terms will have different amplitudes.  In principle, the initial phase at each pulsar is fully determined by the sky location, chirp mass and pulsar distance:
\begin{eqnarray}
\Phi_I  &=& \Phi_0 + \frac{1}{32 {\cal M}^{5/3}}  \left( \omega_0^{-\frac{5}{3}} -  \omega_I^{-\frac{5}{3}} \right) \nonumber \\
&\approx &  \Phi_0 +  (t_0- L_I (1 + \hat{\Omega} \cdot \hat{p}_I)) \, \omega_0 \, .
\label{eq:psrPHASE}
\end{eqnarray}
Small changes in the estimate for the pulsar distance, of order a parsec, cause large changes in $\Phi_I$ but only small changes in $\omega_I$. Since the $\Phi_I$ is only measured modulo $2\pi$, the pulsar distances are in effect only constrained by the frequency measurement. This makes it very difficult to sample the pulsar distance since there are hundreds of local maxima in the likelihood within the envelope of values allowed by frequency measurement. One effective solution is to treat the $\Phi_I$ as independent parameters, which takes care of the phase wrapping problem, at the cost of increasing the size of the parameter space.

The full list of parameters for a single binary black hole is then
\begin{equation}
 \boldsymbol{\zeta} \rightarrow \{\omega_0,   \Phi_0, {\cal M},  d_L, \theta, \phi, \iota, \psi, \Phi_I, L_I\} \, .
\end{equation}
When there are multiple individual binary signals each has its own set of pulsar phase terms, $\Phi_I$, but they all share the same pulsar distance values $L_I$.

The full signal $s(t,  \boldsymbol{\zeta})$ in each pulsar can be expressed in terms of constants amplitudes multiplying the functions $\{\cos(\omega_E t), \sin(\omega_E t)\}$, $\{\cos(\omega_I t), \sin(\omega_I t)\}$, the latter being different for each pulsar in the array. For this reason, individual binary signals are often called continuous wave (CW) signals. The discrete Fourier transform Eq.~(\ref{dft}) can be performed analytically using
\begin{eqnarray}\label{sinc}
\int_0^T \cos(\omega_k t) \cos(\omega t) dt  &=& \frac{ \sin(\omega T)\omega}{\omega^2 - \omega^2_k} \nonumber \\
\int_0^T \sin(\omega_k t) \sin(\omega t) dt  &= &  \frac{ \sin(\omega T)\omega_k}{\omega^2 - \omega^2_k} \nonumber \\
\int_0^T \sin(\omega_k t) \cos(\omega t) dt  &=&  \frac{ (\cos(\omega T)-1)\omega_k}{\omega^2 - \omega^2_k} \nonumber \\
\int_0^T \cos(\omega_k t) \sin(\omega t) dt  &=& -\frac{ (\cos(\omega T)-1)\omega}{\omega^2 - \omega^2_k} \, .
\end{eqnarray}
These expression can be computed cheaply since they only require two evaluations of trigonometric functions, the remaining operations being arithmetic. Moreover, the expressions are highly peaked around $\omega = \omega_k$, so the likelihood can be well approximated using just a few non-zero terms for the Fourier coefficients, ${\bf a}_D$. A numerical Fourier transform may be more efficient, depending on the number of frequency bins used to describe the CW signal.

\subsection{Joint likelihood formulation}
The complete set of model parameters are
\begin{equation}
\boldsymbol{\lambda} = \{  \boldsymbol{\epsilon},  \boldsymbol{\rho},  \boldsymbol{\kappa}, \boldsymbol{\zeta}, \mathbf{a}_R, \mathbf{a}_B\} \, .
\end{equation}
As written, the stochastic components are described by a free spectral model, with no assumption how the power spectrum varies with frequency. Alternatively, the power spectra can be parameterized by a power law with an overall amplitude and spectral index, Eq.~(\ref{eq:powerlaw}). This parameterization can be included as a hyper-prior on the associated Fourier coefficients by parameterizing $\boldsymbol{\rho}$ and $\boldsymbol{\kappa}$.

After specifying priors for the various parameters $\boldsymbol{\lambda}$ the posterior distribution for the model parameters can be found by techniques such as Markov Chain Monte Carlo (MCMC) sampling. However this is not what is usually done since the full parameter space has a very large dimensionality. Instead, by adopting conjugate priors for the parameters $ \{  \boldsymbol{\epsilon},  \boldsymbol{\rho},  \boldsymbol{\kappa} \}$ it is possible to analytically marginalize over the timing model parameters and Fourier coefficients, resulting in a marginalized likelihood function that depends on far fewer parameters, for example, the amplitude and spectral index of the stochastic GWB, and the  amplitude and spectral index of the RN in each pulsar. But there is no free lunch. The marginalization results in a new noise covariance matrix ${\bf K}$ that depends on a complicated combination of the matrices ${\bf N}$, ${\bf M}$ and ${\bf C}$. This dense covariance matrix must be inverted for every posterior evaluation. Crucially, marginalizing over the Fourier coefficients describing the GWB, ${\bf a}_B$, introduces cross terms between pulsars, so the posterior no longer factors per-pulsar. Recently, efficient GPU based implementations have vastly sped up the otherwise very costly matrix operations that result from the marginalization~\cite{GPU_rapid}. These methods use stochastic gradient-descent Bayesian variational inference and rely on a differentiable and parallelized likelihood. Inverting many dense covariance matrices in parallel across a GPU architecture is significantly more efficient than serial inversion required by MCMC methods sampling the marginalized posterior. In this paper, we avoid dense matrix inversions in general by directly sampling the Fourier coefficients. Unless the white noise model is updated, no matrix inversions are required in the hyper-efficient likelihood evaluations making MCMC sampling a viable approach.

In our approach we only analytically marginalize over the timing model, which amounts to replacing ${\bf N}^{-1}$ in the likelihood by $\Tilde{{\bf N}}^{-1} \equiv {\bf G}({\bf G}^T {\bf N} {\bf G})^{-1} {\bf G}^T$ where the matrix ${\bf G}$ is built from the design matrix ${\bf M}$ as in Ref.~\cite{Gmatrix, Gmatrix1, Gmatrix2}. We'll adopt a shared Fourier basis for the background and red noise in our analysis, $\mathbf{F}_B = \mathbf{F}_R \equiv \mathbf{F}$, and use one set of combined coefficients $\mathbf{a}_{R + B} = \mathbf{a}_R + \mathbf{a}_B$ to describe the stochastic processes. The combined set of Fourier coefficients is modeled by a zero mean, Gaussian distribution with covariance matrix given by Eq.~(\ref{combined-cov}). A distinct Fourier basis with an extended period is used to describe the CW model $\mathbf{F}_D\neq\mathbf{F}$ as discussed in Section~\ref{sec:DetSigFR} and Appendix~\ref{app:F_CW}. The Fourier coefficients corresponding to the CW signal, $\mathbf{a}_D$, are calculated via the FFT of the CW model. That is, there is a deterministic mapping from the CW parameters, $\boldsymbol{\zeta}$, to the coefficients, $\mathbf{a}_D = \mathbf{a}_D(\boldsymbol{\zeta})$.

The log-likelihood (up to an additive constant) is
\begin{widetext}
\begin{equation}
\begin{split}
\ln p(\delta {\bf t}|\mathbf{a}_{R+B}, \boldsymbol{\zeta}) &=  -\frac{1}{2}\bigg[{\bf \delta t} - \mathbf{F}\,\mathbf{a}_{R + B} - \mathbf{F}_D\,\mathbf{a}_D(\boldsymbol{\zeta})\bigg]^\text{T}\,\tilde{\mathbf{N}}^{-1}\bigg[{\bf \delta t} - \mathbf{F}\,\mathbf{a}_{R + B} - \mathbf{F}_D\,\mathbf{a}_D(\boldsymbol{\zeta})\bigg] \\
&=-\frac{1}{2} \bigg[U - 2\,{\bf V}^\text{T}\,{\bf a}_{R+B} - 2\mathbf{V}^\text{T}_D\,\mathbf{a}_D(\boldsymbol{\zeta}) + {\bf a}_{R+B}^\text{T}\,{\bf W}\,{\bf a}_{R+B} \\
&\hspace{19mm}+\mathbf{a}_D(\boldsymbol{\zeta})^\text{T}\,\mathbf{W}_D\,\mathbf{a}_D(\boldsymbol{\zeta}) + 2\mathbf{a}^\text{T}_{R+B}\,\tilde{\mathbf{W}}\,\mathbf{a}_D(\boldsymbol{\zeta})\bigg]
\end{split}
\end{equation}
\end{widetext}
where $U$ is a scalar, $\mathbf{V}$ and $\mathbf{V}_D$ are vectors, and $\mathbf{W}$, $\mathbf{W}_D$, and $\tilde{\mathbf{W}}$ are matrices. They are defined using the inner product $(\mathbf{u}|\mathbf{v}) = \mathbf{u}^\text{T}\tilde{\mathbf{N}}^{-1}\mathbf{v}$ as

\begin{eqnarray}\label{eq:inner_products}
U &=& ( \delta {\bf t}   | \delta {\bf t}  )\nonumber \\
{\bf V}^\text{T} &=& ( \delta {\bf t}   | {\bf F} )\nonumber \\
{\bf V}_D^\text{T} &=& ( \delta {\bf t}   | {\bf F}_D )\nonumber \\
{\bf W} &=& ({\bf F}   | {\bf F} )\nonumber \\
{\bf W}_D &=& ({\bf F}_D   | {\bf F}_D )\nonumber \\
\tilde{\mathbf{W}} &=& ({\bf F}   | {\bf F}_D )\,.
\end{eqnarray}

So long the white noise model is held fixed these inner products can be computed once and stored. Alternatively, the inner products can be updated periodically if the white noise model is updated in a blocked sampling scheme. The ${\bf W}_{(D)}$ matrix would be diagonal if the data were evenly sampled across the full observation span, but uneven sampling leads to off-diagonal terms. However, the matrix is still diagonal dominant, especially at high frequencies, and most of the off-diagonal terms can be set to zero. We choose to keep all elements of the ${\bf W}_{(D)}$ matrix in this paper, rather than setting off-diagonal elements to zero. Keeping the white noise model fixed, the matrix operations in our likelihood scale as $\mathcal{O}(4N_p^2N_f^2)$ where $N_p$ is the number of pulsars in the array and $N_f$ the number of frequency bins modeled. The matrix operations in the likelihood scale as $\mathcal{O}(2N_pN_f)$ if we neglect all off-diagonal elements of the ${\bf W}_{(D)}$ matrix. In practice, the off-diagonal elements at low frequencies are significant and must be kept in the matrix products. Moreover, the $\tilde{\mathbf{W}}$ matrix can have significant entries in the band diagonal.

\section{Prior and Posterior Formulation}\label{sec:priors}

The priors on the RN and GWB Fourier coefficients take the form of a zero mean multi-variate Gaussian distribution with a covariance matrix given by Eq.~(\ref{cov}) or Eq.~(\ref{combined-cov}); which covariance matrix is used depends on whether the GWB and pulsar RN are modeled with a distinct or combined set of coefficients. We find the sampling converges more rapidly when one set of coefficients, $\mathbf{a}_{R+B}$, model both the background and pulsar red noise. If distinct sets of coefficients, $\mathbf{a}_R$ and $\mathbf{a}_B$, are used to model the red noise and background respectively, the number of parameters in the likelihood nearly double. This greatly increases the parameter volume to be explored by the sampler. Besides the increase in dimension, sampling two sets of coefficients is challenging because the stochastic signals they represent are nearly degenerate. Both stochastic processes are well-approximated by a power law and the modeling degeneracy is only broken via the inter-pulsar Hellings-Downs correlation pattern, Eq.~(\ref{eq:HD-corr}). While there is no physical correlation between the two stochastic processes, their respective parameters become correlated during the sampling because of this modeling similarity. To ease the difficulty of sampling correlated high-dimensional parameter spaces, the GWB and pulsar RN are modeled with one combined set of Fourier coefficients.

There are a total of $2 N_p N_f$ Fourier coefficients for the stochastic processes, where $N_p$ is the number of pulsars and $N_f$ the number of frequency bins in the model. We may directly sample these coefficients using a free spectral model where the power at each frequency, $\{ \rho_i, \kappa_{Ii}\}$ are allowed to take on any value. A particular spectral model can then be applied in post-processing, see Appendix~\ref{app:NealsFunnel}. Alternatively, the spectral model can be applied as a hyper-prior, with the parameters of the spectral model (e.g. amplitude and spectral index) as hyper-parameters. While it can be more efficient to impose the spectral models directly, the sampling can get trapped in ``Neal's funnel''~\cite{neal2003slice} unless appropriate reparameterizations or sampling strategies are applied~\cite{reparameterization}. For further discussion on Neal's funnel in the context of PTA datasets, see Appendix~\ref{app:NealsFunnel}. In our analysis, the pulsar RN will be modeled with a power law hyper-prior. The GWB will be modeled using both a free spectral and power law hyper-model (independently) for comparison.

At a given frequency, if the power spectrum model has amplitude $\beta_k$ then the prior on the sine and cosine amplitudes  $\{a^s_k, a^c_k\}$ each follow a Gaussian distribution with zero mean and variance $\beta^2_k/2$. Alternatively, we can use a polar parameterization and impose a prior on polar coordinates $\{A_k, \phi_k\}$ where $a^s_k = A_k \sin \phi_k$, $a^c_k = A_k \cos \phi_k$. The $\phi_k$ follow a uniform distribution in the range $[0,2\pi)$, while the scaled amplitudes $A^2_k/\beta^2_k$ follow a standard chi-squared distribution with two degrees of freedom. In our current implementation we chose to model the (Cartesian) sine and cosine amplitudes, $\{a^s_k, a^c_k\}$.

For CW signals originating from individual SMBHBs we use priors that are uniform in $\{\log_{10}\omega_0, \Phi_0, \log_{10} {\cal M},  \psi, \cos\iota, \Phi_L\}$ and uniform in space for $\{\log_{10}d_L, \cos\theta, \phi\}$. For the pulsar distances, we use Gaussian priors on $L_I$, centered on the best estimate found from the dispersion measure (and parallax if available), with a variance estimated from those measurement techniques.

Up to an overall additive constant, the log-posterior is given by

\begin{widetext}
\begin{equation}\label{eq:posterior}
\begin{split}
\ln p(\boldsymbol{\rho}, \boldsymbol{\kappa}, \boldsymbol{\zeta}, \mathbf{a}_{R+B}|\delta {\bf t}) &=-\frac{1}{2} \bigg[U - 2\,{\bf V}^\text{T}\,{\bf a}_{R+B} - 2\mathbf{V}^\text{T}_D\,\mathbf{a}_D(\boldsymbol{\zeta}) + {\bf a}_{R+B}^\text{T}\,{\bf W}\,{\bf a}_{R+B} \\
&\hspace{40mm}+\mathbf{a}_D(\boldsymbol{\zeta})^\text{T}\,\mathbf{W}_D\,\mathbf{a}_D(\boldsymbol{\zeta}) + 2\mathbf{a}^\text{T}_{R+B}\,\tilde{\mathbf{W}}\,\mathbf{a}_D(\boldsymbol{\zeta}) \\
&\hspace{40mm}+ \mathbf{a}_{R + B}^\text{T}\,\mathbf{C}_{R+B}^{-1}\,\mathbf{a}_{R+B} + \ln\text{det}(2\pi\mathbf{C}_{R+B})\bigg]
\end{split}
\end{equation}
\end{widetext}
where we choose to describe the GWB and RN with one set of Fourier coefficients, and the joint covariance matrix in the hyper-prior is given by Eq.~(\ref{combined-cov}). The power spectrum for the pulsar RN, $\boldsymbol{\kappa}_I$, is parameterized with a power law, Eq.~(\ref{eq:powerlaw}), and the power spectrum for the GWB, $\boldsymbol{\rho}$, uses a free spectral model and (independently) a power law parameterization for comparison.

\section{Sampling strategies}

The posterior density, Eq.~(\ref{eq:posterior}), is implemented with the JAX~\cite{JAX} package and therefore utilizes automatic differentiation. The model parameters can be efficiently sampled using a Hamitlonian Monte Carlo (HMC) No U-Turn Sampler (NUTS) from the \texttt{NumPyro} package~\cite{NumPyro1, NumPyro2}. We also find a parallel tempered Markov Chain Monte Carlo (PTMCMC) algorithm is able to sample the large parameter space. 

To achieve rapid convergence, we need well-tailored proposal distributions. We find a blocked sampling scheme which updates each signal and noise process in turn eases the difficulty of sampling the high-dimensional parameter space with PTMCMC. Some proposals are made with differential evolution~\cite{DiffEvol1, DiffEvol2} to more rapidly resolve correlations in the posterior.

Another proposal distribution we have found to work well in a range of settings are jumps along eigenvectors of the augmented Fisher information matrix $\boldsymbol{\Sigma}$, with the jump sizes scaled by the inverse square root of the eigenvalues. The augmented Fisher matrix is given by the negative Hessian of second derivatives of the log posterior, meaning that it includes the effects of the priors. For parameters that are well constrained by the likelihood the augmented Fisher matrix is dominated by the contribution from ordinary Fisher matrix, while for parameters that are poorly constrained, using the augmented Fisher matrix is equivalent to making draws from the prior.

If the GWB and RN are modeled with distinct sets of Fourier coefficients, $\mathbf{a}_B$ and $\mathbf{a}_R$, the ordinary Fisher matrix is singular due to degeneracies between coefficients in the likelihood. Including the priors in the augmented Fisher matrix breaks this degeneracy. The stochastic components of the augmented Fisher matrix are given by
\begin{eqnarray}
\Sigma_{a^B_{Ii} a^B_{Jj}} &=& W_{Ii ,Jj} +  C^{-1}_{(B)\,\,Ii,Jj}  \nonumber \\
\Sigma_{a^B_{Ii} a^R_{Jj}} &=& W_{Ii ,Jj}   \nonumber \\
\Sigma_{a^R_{Ii} a^R_{Jj}} &=& W_{Ii ,Jj} +  C^{-1}_{(R)\,\,Ii,Jj} \, .
\end{eqnarray}
Notice $\Sigma_{a^B a^R}\neq0$, illustrating the covariance between the two stochastic processes. If instead we model both stochastic processes with one set of coefficients, the elements of the augmented Fisher matrix are

\begin{equation}
    \Sigma_{a^{R+B}_{Ii}a^{R+B}_{Jj}} = W_{Ii,Jj} + C^{-1}_{(R+B)\,\, \\ Ii, Jj}\,.
\end{equation}
Using Greek letters to denote the parameters of the deterministic model, the component of the augmented Fisher matrix for the CW has the form
\begin{align}
\Sigma&_{\alpha \beta}
   = \delta_\alpha^{L_I} \delta_\beta^{L_J} \delta_{IJ} 
      \,\frac{1}{\sigma^2_{L_I}}
      + W^D_{Ii ,Jj} 
      \,\dfrac{\partial a^D_{Ii}}{\partial \zeta^\alpha}  
      \dfrac{\partial a^D_{Jj}}{\partial \zeta^\beta} \nonumber \\
   &\quad+ \bigg(
      W^D_{Ii, Jj}\,a^D_{Ii}  
      + \tilde{W}_{Ii, Jj}\,a^{R+B}_{Ii} - V^D_{Jj}
      \bigg)
      \dfrac{\partial^2 a^D_{Jj}}
            {\partial\zeta^\alpha\,\partial\zeta^\beta}
\end{align}
In most instances, the terms in the augmented Fisher matrix that involve the pulsar distances are dominated by the contribution from the prior. It is important not to forget the cross terms between the stochastic models and the deterministic model. For example,
\begin{equation}
\Sigma_{\alpha a^B_{Ii}} = \tilde{W}_{Ii ,Jj} \frac{\partial a^D_{Jj}}{\partial \zeta^\alpha}  \, .
\end{equation}

In practice the augmented Fisher matrix is updated periodically as we sample, evaluating the Fisher matrix at the current state of the MCMC chain. Additionally, analytic \textit{maximum a posteriori} (MAP) solutions for the Fourier coefficients can be used to make the Fisher proposals more effective. At some iteration of the MCMC the hyper-model and CW parameters are at values $(\hat{\rho}, \hat{\kappa}, \hat{\zeta})$ and the (conditional) MAP Fourier coefficients $\mathbf{\hat{a}}_{R+B}$ are determined by the set of equations $\partial_{\mathbf{a}_{R + B}} \ln p(\boldsymbol{\lambda} | {\bf d})\vert_{\hat{\boldsymbol{\lambda}}} = 0$, whose solution is
\begin{eqnarray}
\hat{\mathbf{a}}_{R+B} = \big[\mathbf{W} + \mathbf{C}_{R+B}^{-1}(\hat{\rho}, \hat{\kappa})\big]^{-1}\,\big[\mathbf{V} - \tilde{\mathbf{W}}\,\mathbf{a}_D(\hat{\boldsymbol{\zeta}})\big]\,.
\end{eqnarray}
We can then evaluate the Fisher matrix at this MAP solution, and propose Fisher jumps about such maxima.

\section{Results}\label{Results}

\subsection{Simulated dataset}

To assess the capabilities of this joint model, we use the methods above to analyze simulated datasets. More complete noise modeling~\cite{NGnoisebudget} is required before these methods can be applied to real datasets. Twenty pulsars are simulated and their timing residuals are generated consistent with the signal and noise models discussed in Section~\ref{model}. The sky location of each pulsar is drawn isotropically from the sphere, and their distance from a uniform distribution, $L_I\sim \text{Uniform}[0.1,\,6]\;(\text{kpc})$. Every pulsar in the simulated dataset is observed monthly for 15 years. As PTA data analysis is characterized by unevenly spaced samples in the time-domain, we randomly offset each TOA with a zero mean Gaussian draw with a standard deviation of 2 days. The TOA measurement uncertainty is fixed at $0.5\,\mu\text{s}$ for every observation. 

Generally NANOGrav models three kinds of white noise: EFAC, ECORR, and EQUAD~\cite{NGnoisebudget}. To simplify the analysis, we neither simulate nor model ECORR or EQUAD. We simulate EFAC=1 in every pulsar, and fix the white noise model over the analysis. We inject intrinsic pulsar RN obeying a power law in each of the simulated pulsars. The injected amplitudes of the power law are drawn from the log-uniform distribution, $\log_{10}A_R\sim \text{Uniform}[-18, -14]$, and spectral indices are drawn from $\gamma_R\sim \text{Uniform}[2, 7]$. We inject a stochastic GWB obeying the Hellings-Downs correlation using a power law with hyper-parameters $\log_{10}A_\text{B} = -14$ and $\gamma_\text{B} = 13/3$. The background and intrinsic red noise are simulated and modeled using $N_f=10$ frequency bins.

A single deterministic CW signal with parameters $\omega_0 \equiv 2\pi f_\text{GW} = 2\pi(4\times 10^{-9}\,\text{[Hz]})$, $\Phi_0 = 0$, $\log_{10}(\mathcal{M} / \,[\text{M}_\odot]) = 8.6$, $d_L = 1\,\text{Mpc}$, $\theta = 2\pi/5$, $\phi = 7\pi/4$, $\iota = 0$, and $\psi=0$ is also injected. The pulsar phases are determined by Eq.~(\ref{eq:psrPHASE}). The particular injections above yield a CW SNR of 13.7. Although the methods presented in this paper are amenable to any arbitrary set of deterministic models, we inject and model only one CW signal. A simple quadratic timing model is fit to the simulated dataset (accounting for spin period, period derivative, and phase offset). In the analysis, the parameters of the timing model are analytically marginalized out of the posterior density by projecting the data into a space orthogonal to the timing model as in Refs.~\cite{Gmatrix, Gmatrix1, Gmatrix2}.

\subsection{Results on simulated data}
\begin{figure*}
    \centering
    \includegraphics[width=0.8\linewidth]{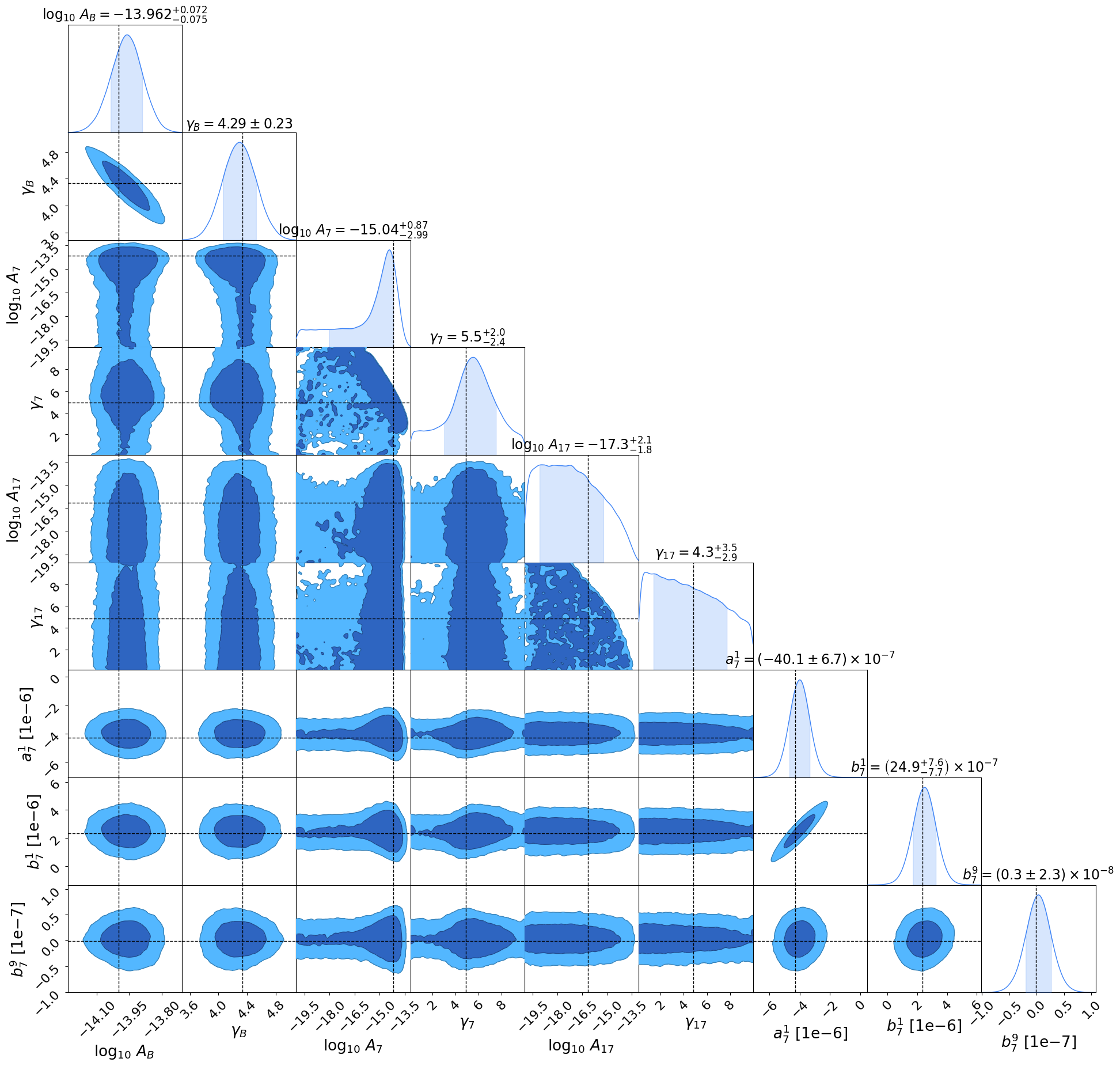}
    \caption{Samples of selected parameters for the stochastic processes obtained via HMC. The first two columns correspond to the power law hyper-parameters of the GWB, and subsequent columns are power law hyper-parameters for intrinsic pulsar RN and the Fourier coefficients. The ``$a$" Fourier coefficients scale sine functions and ``$b$" scale cosine. Parameter subscripts index the pulsars in the array, superscripts on the Fourier coefficients index the corresponding frequency bins. The dashed lines are the parameter values injected into the simulated data. The summary statistics and shading of one-dimensional marginal distributions is $1\sigma$ either side the median computed with the CDF. Contours of the two-dimensional distributions enclose the $1\sigma$ and $2\sigma$ credible regions.}
    \label{fig:GWBcorner}
\end{figure*}

We sample the posterior probability density Eq.~(\ref{eq:posterior}) over the model parameters using a Hamiltonian Monte Carlo (HMC) with the \texttt{NumPyro} package and its No U-Turn Sampler (NUTS)~\cite{NumPyro1, NumPyro2}. One set of Fourier coefficients is used to model both the background and red noise. The GWB is modeled using a power law and (independently) a free spectral model for comparison. The pulsar RN is modeled with a power law. The recovered posterior on a subset of parameters describing the stochastic processes is shown in Figure~(\ref{fig:GWBcorner}). The injected parameter values all lie within the posterior volume. Using a reference frequency $f_\text{ref}=1\,\text{yr}^{-1}$, the expected correlation between amplitude and spectral index for a power law is recovered.

\begin{figure*}
    \centering
    \includegraphics[width=0.8\linewidth]{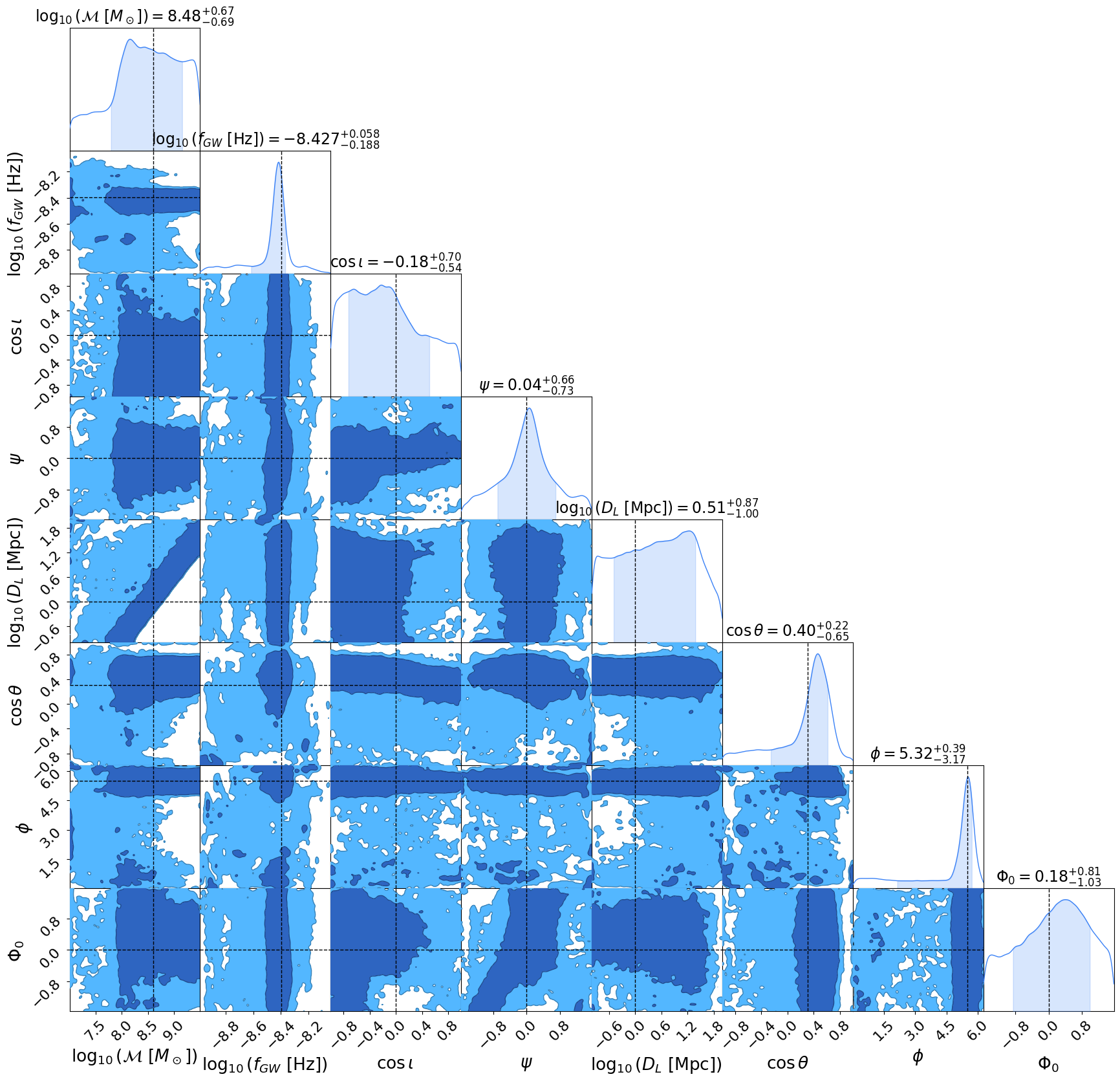}
    \caption{Samples of continuous wave model parameters obtained via HMC. The CW SNR is 13.7. The dashed lines are the parameter values injected into the simulated dataset. The summary statistics and shading of one-dimensional marginal distributions is $1\sigma$ either side the median computed with the CDF. Contours of the two-dimensional distributions enclose the $1\sigma$ and $2\sigma$ credible regions.}
    \label{fig:CWcorner}
\end{figure*}

\begin{figure*}
    \centering
    \includegraphics[width=0.8\linewidth]{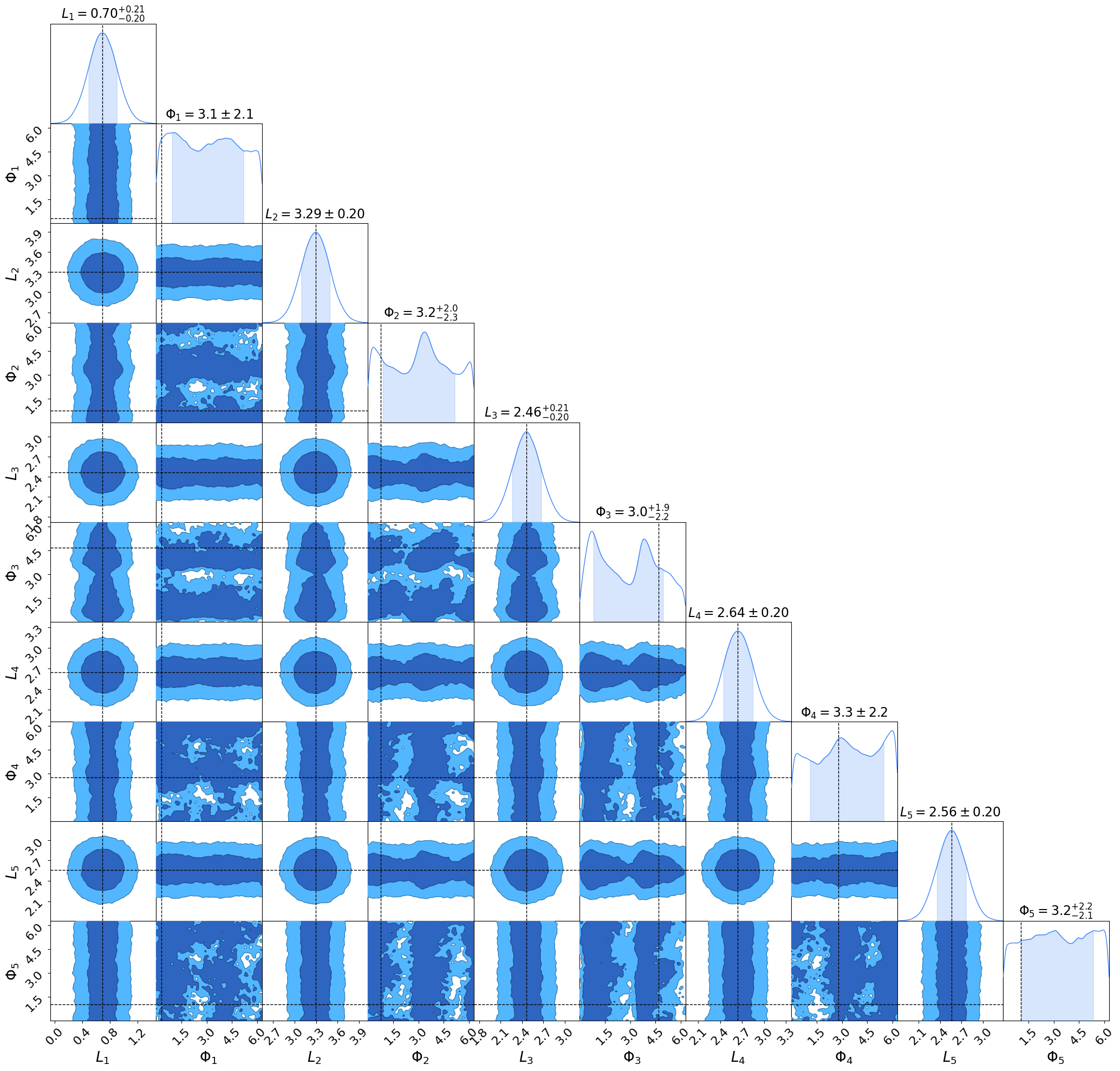}
    \caption{Samples of selected pulsar distance and phase parameters obtained via HMC. Parameter subscripts index pulsars in the array. The dashed lines are the parameter values injected into the simulated dataset. The summary statistics and shading of one-dimensional marginal distributions is $1\sigma$ either side the median computed with the CDF. Contours of the two-dimensional distributions enclose the $1\sigma$ and $2\sigma$ credible regions.}
    \label{fig:Psrcorner}
\end{figure*}

\begin{figure*}
    \centering
    \includegraphics[width=0.8\linewidth]{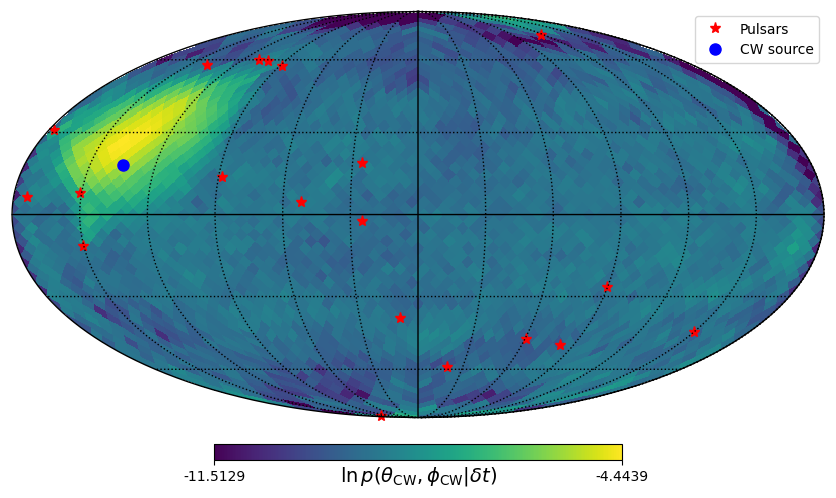}
    \caption{Sky map illustrating resolvability of the CW source's sky location. The red stars are pulsars observed in the PTA. The blue circle is the CW source. The color of the sky map is the log-posterior density (numerically) marginalized over all parameters except sky location.}
    \label{fig:skymap}
\end{figure*}

The posterior on the CW parameters is shown in Figure~(\ref{fig:CWcorner}), and subset of recovered pulsar distances and phases (used in the CW model) are shown in Figure~(\ref{fig:Psrcorner}). A sky map in Figure~(\ref{fig:skymap}) illustrates resolvability of the CW source. Again all injected parameter values are found within the spread of the posterior.

\begin{figure}
    \centering
    \includegraphics[width=0.95\linewidth]{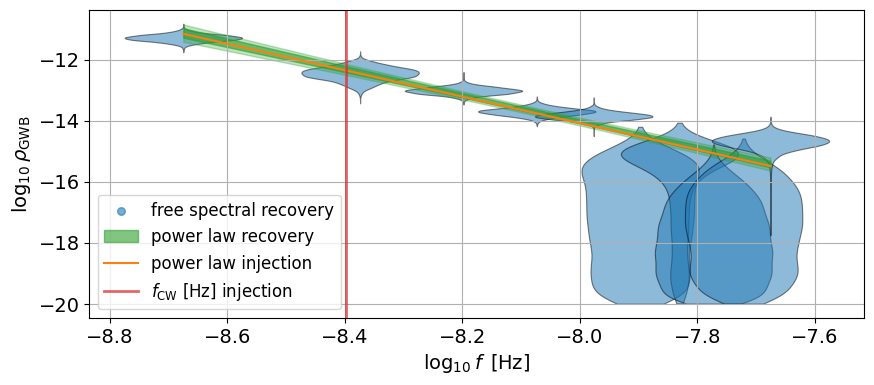}
    \caption{Violin plot illustrating the recovered power for the stochastic GWB from a free spectral run. The injected GWB power law is shown in orange. The dark (light) green shading corresponds to a $1\sigma$ ($2\sigma$) uncertainty bound in the hyper-parameters from a hierarchical power law analysis on the same dataset. The vertical red line is the frequency of the injected CW signal.}
    \label{fig:violin}
\end{figure}

The recovered power of the stochastic GWB from a free spectral analysis is shown in Figure~(\ref{fig:violin}). The spectrum recovered from a hierarchical power law run is consistent with the free spectral analysis. Both analyses use the same simulated dataset and model, only differing in the parameterization of the covariance matrix for the Fourier coefficients, Eq.~(\ref{combined-cov}). The injection of a CW signal does not bias the recovery of the power law or free spectral model. That is, there is no excess power in frequency bins describing the stochastic processes near the CW signal's frequency. The joint analysis models the CW while simultaneously fitting the spectra of red processes.

\begin{figure}
    \centering
    \includegraphics[width=0.95\linewidth]{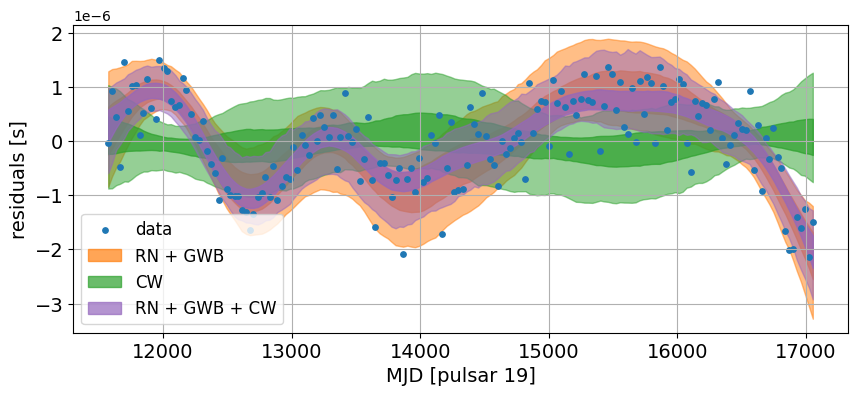}
    \caption{Bayesogram showing the time-domain recovery of the signal and noise processes in a particular pulsar. The dark (light) shading corresponds to a $1\sigma$ ($2\sigma$) uncertainty bound.}
    \label{fig:Bayes}
\end{figure}

Figure~(\ref{fig:Bayes}) shows the time-domain recovery of the CW signal and the combined stochastic GWB and pulsar RN processes. While the various signals and noise were simulated and injected independently, there is observable correlation in the recovery of the models. While sampling the posterior, certain sets of parameters allow the CW model to ``absorb" some power originating from the red processes and vice versa. In other words, the uncertainty in the individual signal and noise recovery is larger than the uncertainty of their sum, indicating a sampling covariance.

\section{Future work}

Before this method is applied to real datasets, more complete noise modeling must be incorporated~\cite{NGnoisebudget}. This can be accomplished with a blocked sampling scheme, where the white noise model is occasionally updated, and the inner products in the likelihood are recomputed and stored for the next set of sampling. Additionally, real datasets require increased frequency resolution in modeling the intrinsic pulsar RN. Neal's funnel arises in the posterior when more frequency bins are modeled, see Appendix~\ref{app:NealsFunnel}. Therefore, a more efficient sampling scheme for the Fourier coefficients is required before these methods can be applied to real datasets. We are currently in collaboration with a subset of authors from Refs.~\cite{Laal:2023etp, Laal:2024hdc} for this purpose and the software will be released after further development and testing. The CW modeling techniques developed in this paper will be combined with the Gibbs sampling scheme for the Fourier coefficients as described by the previous references.

A model which is trans-dimensional in the number of Fourier coefficients will also be developed. As the coefficients describe red processes, the information each coefficient contributes to the likelihood decreases as the frequency bins increase. After a certain bin, the Fourier coefficients simply recover their prior. Reversible-jump MCMC (RJMCMC) allow us to sample models with different numbers of Fourier modes, and determine the number of coefficients supported by the data. A trans-dimensional model would also allow the simultaneous modeling of various sets of deterministic signals. For example, RJMCMC could move between models with varying numbers of CW signals~\cite{Becsy:2019dim}. The same techniques used here for CW signals can be used to search for other deterministic signals such as gravitational wave bursts~\cite{Becsy:2020utk}. We will also look at using our Fourier domain CW model with the per-pulsar Bayesian Fourier transform posteriors described in Ref.~\cite{RF}.

\section*{Acknowledgments}
This project was supported by National Science Foundation (NSF) Physics Frontiers Center award 2020265. We thank Bence B\'ecsy and Nima Laal for valuable discussions. We'd also like to thank Rutger van Haasteren for suggestions and example code which samples the Fourier coefficients in a Hamiltonian Monte Carlo scheme. The code used to simulate data, run analyses, and produce figures presented in this paper is publicly available at~\cite{code}.

\appendix

\section{Fourier representation of continuous waves}\label{app:F_CW}

As discussed in Section~\ref{sec:DetSigFR}, deterministic signals need not be represented in a Fourier basis. It's possible to extend the treatment of Ref.~\cite{Lentati:2012xb} to a joint analysis by simply evaluating a deterministic model over the observed TOAs, and combining the induced timing delays linearly with those of the stochastic processes. However if the deterministic model is expensive to evaluate and can be described accurately with relatively few frequency bins, as is the case of CW signals from individual SMBHBs, then a Fourier representation is a computationally efficient method to obtain the deterministic model over the observed TOAs.

Rather than evaluating the CW model over the observed TOAs, we evaluate the model over a relatively sparse collection of evenly spaced time samples $\mathbf{t}_m$. Applying a fast Fourier transform (FFT) gives us the associated Fourier coefficients, $\mathbf{a}_D$, for the particular realization of the deterministic model. Then applying the Fourier design matrix $\mathbf{F}_D$, defined from the extent and density of $\mathbf{t}_m$, to these coefficients recovers the deterministic model evaluated over the observed TOAs. This method is similar to interpolation, where we start with the deterministic model evaluated on $\mathbf{t}_m$ and interpolate to the observed TOAs. The interpolation method here is the composition of the Fourier transform and design matrix.

The FFT assumes the deterministic time-domain signals are periodic over the observation window, which is generally not true. The time-domain deterministic signal reconstructed from the Fourier basis therefore exhibits Gibbs phenomena at either end of the sampling window~\cite{gibbs}. To remedy this, we extend the sampling window for $\mathbf{t}_m$ either side of the observation window. Applying a Tukey window that is flat over the observation span and tapered in the extended regions further reduces spectral leakage~\cite{1978IEEEP..66...51H}. Altogether, the density of $\mathbf{t}_m$ must be chosen to resolve all desired frequencies in the deterministic signal. The extent of $\mathbf{t}_m$ must be chosen to reduce Gibbs phenomena in the reconstructed signal. Below we test various choices for $\mathbf{t}_m$, and evaluate the accuracy and computational efficiency of the Fourier representation for CW signals.

Let $\mathbf{\tilde{t}}_\text{D}$ denote the CW model evaluated over the observed TOAs and $\boldsymbol{\widetilde{(F a)}}_D$ the reconstruction of the CW signal using a Fourier representation. As we analytically marginalize over a quadratic timing model in the analysis above, we project both signals into a space orthogonal to the timing model. We assess the accuracy of the Fourier representation by computing the power in the relative error between models over various choices for $\mathbf{t}_m$,
\begin{equation}
    \delta = \frac{|\mathbf{\tilde{t}}_\text{D} - \boldsymbol{\widetilde{(F a)}}_D|^2}{|\mathbf{\tilde{t}}_\text{D}|^2}
\end{equation}\label{rel_error}

\begin{figure}[h]
    \centering
    \includegraphics[width=0.95\linewidth]{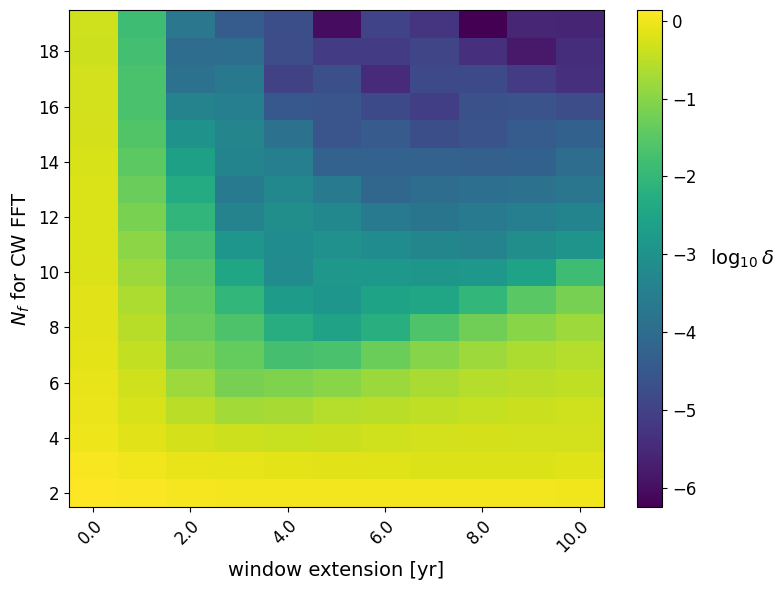}
    \caption{The power in the relative error between the two CW waveform models in a 20 pulsar PTA: the deterministic CW model evaluated over observed TOAs and the Fourier reconstructed signal. The relative error is averaged over 10,000 random CW signals drawn from their prior distribution. The window extension measures how far either side of the observation window to sample $\mathbf{t}_m$ (reducing Gibbs phenomena). $N_f$ is the maximum frequency bin resolved by the FFT and is related to the density of $\mathbf{t}_m$.}
    \label{fig:CW_test}
\end{figure}

As discussed above, the sampling window for $\mathbf{t}_m$ is extended either side of the observation window to reduce Gibbs phenomena, and the density of samples dictates the highest frequency bin resolved. The mean power of the relative error between the CW model and its Fourier reconstruction is computed for 10,000 random CW signals (drawn from their prior). The power in the relative error is plotted as a function of window extension and the highest frequency bin resolved from the FFT in Figure.~(\ref{fig:CW_test}). If the window extension is sufficiently long and $\mathbf{t}_m$ samples sufficiently dense, then the Fourier reconstruction is an excellent approximation to the CW model. For example, using a window  years (either side) and modeling up to $N_f = 15$ frequency bins, the mean power in the relative error between models is approximately $0.002\%$, and the maximum relative error out of 10,000 random CW signals is less than $0.2\%$. On a personal laptop, it takes over twice as long to evaluate the CW model over the observed TOAs than reconstructing the signal from a Fourier basis.

\section{One pulsar analysis and Neal's funnel}\label{app:NealsFunnel}

Neal's funnel is a funnel-like posterior geometry from which it's difficult to sample, and a common feature of Bayesian hierarchical models~\cite{neal2003slice}. It occurs when a hyper-parameter controls the variance of a latent parameter in the prior. When the variance of the latent parameter is small, a narrow but dense probability region forms the neck of the funnel. The opening of the funnel is formed where the variance of of the latent parameter is large. The neck of the funnel is difficult to sample because precise jump proposals must be made and adapted along the funnel. Naive samplers often get stuck or fail to explore the full posterior, resulting in poor convergence for such hierarchical models.

In PTA datasets, Neal's funnel arises when sampling Fourier coefficients which describe a red process with a hierarchical power law imposed in the prior. The amplitude and spectral index of the power law constrain the power in each frequency bin, and the conditional distribution on Fourier coefficients can have wide range of variance, resulting in Neal's funnel. We illustrate this funnel geometry with a red noise analysis in one pulsar. A single red noise process is simulated in a pulsar with a set of hyper-parameters, amplitude and spectral index ($\log_{10}A$ and $\gamma$). Simulated white noise is drawn from a zero mean Gaussian distribution with fixed variance, and we project the data into a space orthogonal to a quadratic timing model. We simulate data using $N_f=30$ frequency bins.

\subsection{Hierarcical power law modeling in a single pulsar} \label{one_psr_power}

We sample the posterior density Eq.~(\ref{eq:posterior}), where the data consists only of red and white noise in a single pulsar, and the model parameters are the Fourier coefficients which describe the red process $\mathbf{a}_R$ and their hyper-parameters $\log_{10}A$ and $\gamma$. The white noise parameters are fixed at their injected values. Neither a gravitational wave background nor continuous wave is injected into the data, and we neglect their contributions to the posterior. The posterior density is sampled using a Hamiltonian Monte Carlo (HMC) from the \texttt{NumPyro} package and its No U-Turn Sampler (NUTS)~\cite{NumPyro1, NumPyro2}.

\begin{figure*}
    \centering
    \includegraphics[width=0.8\linewidth]{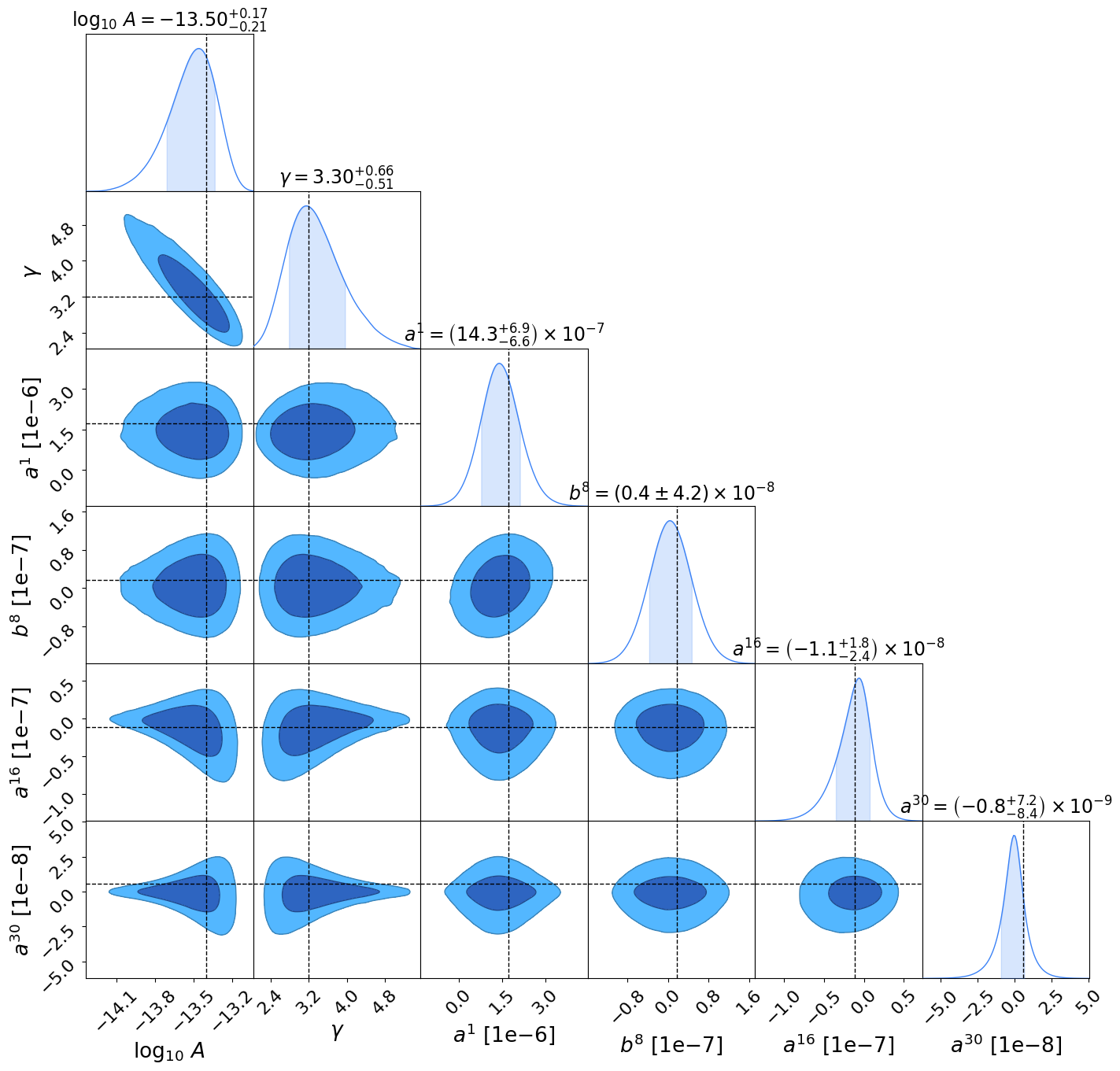}
    \caption{Samples of selected parameters obtained via HMC in a hierarchical power law red noise analysis for a single pulsar. The first two columns correspond to the hyper-parameters of the power law. The remaining columns are samples of the Fourier coefficients. The ``$a$" Fourier coefficients scale sine functions and ``$b$" scale cosine. Parameter superscripts index frequency bins. Neal's funnel is observed in the posterior samples plotted over the hyper-parameters and the Fourier coefficients in higher frequency bins. The dashed lines are the parameter values injected into the simulated data. The summary statistics and shading of one-dimensional marginal distributions is $1\sigma$ either side the median computed with the CDF. Contours of the two-dimensional distributions enclose the $1\sigma$ and $2\sigma$ credible regions.}
    \label{fig:NealsCorner}
\end{figure*}

The corner plot for a selection of parameters is shown in Figure~(\ref{fig:NealsCorner}). Neal's funnel is observed in the plot between hyper-parameters and the Fourier coefficients for high frequency bins. Because the data is simulated according to a power law, there is relatively low power in high frequency bins and the marginal distribution on respective Fourier coefficients is centered near zero. The variance of the distribution of Fourier coefficients is small when the amplitude (spectral index) of the power law is small (large). Conversely, the variance Fourier coefficients in high frequency bins is large when the amplitude (spectral index) of the power law is large (small).

There are various ways to remedy sampling issues due to Neal's funnel. In standard PTA data analysis, the Fourier coefficients are regarded as nuisance parameters and analytically marginalized out of the posterior. As discussed above, this yields a dense covariance matrix which must be inverted in every posterior evaluation making the analysis computationally expensive. However, the complex geometry of Neal's funnel is marginalized over with this approach and a simpler posterior geometry remains.

Alternatively, one may reparameterize the distribution to escape Neal's funnel~\cite{reparameterization}. Samplers which fail to converge on distributions exhibiting Neal's funnel may become well-behaved after a clever reparameterization, or obtain effective samples at a higher rate. Another option is to simply use a more robust sampler to handle Neal's funnel: HMC~\cite{HMCHierarch} and its extension, Riemannian Manifold Hamiltonian Monte Carlo (RMHMC)~\cite{RMHMC1, RMHCM2} which uses a position-dependent metric, have proven effective at sampling Neal's funnel in hierarchical models. In PTA datasets, Gibbs sampling has also been shown an effective method for sampling the Fourier coefficients~\cite{Laal:2023etp, Laal:2024hdc}.

\subsection{Free spectral modeling in a single pulsar}

A potential way to reduce the difficulty in sampling Neal's funnel, in the case of PTA data analysis, is to describe the red process using a free spectral model. That is, instead of parameterizing the covariance matrix for the Fourier coefficients Eq.~(\ref{cov}) with power law hyper-parameters (amplitude and spectral index), every diagonal element is itself a free parameter. These free parameters describe the power in each frequency bin due to the red process and can be sampled with log-uniform priors.

\begin{figure*}
    \centering
    \includegraphics[width=0.8\linewidth]{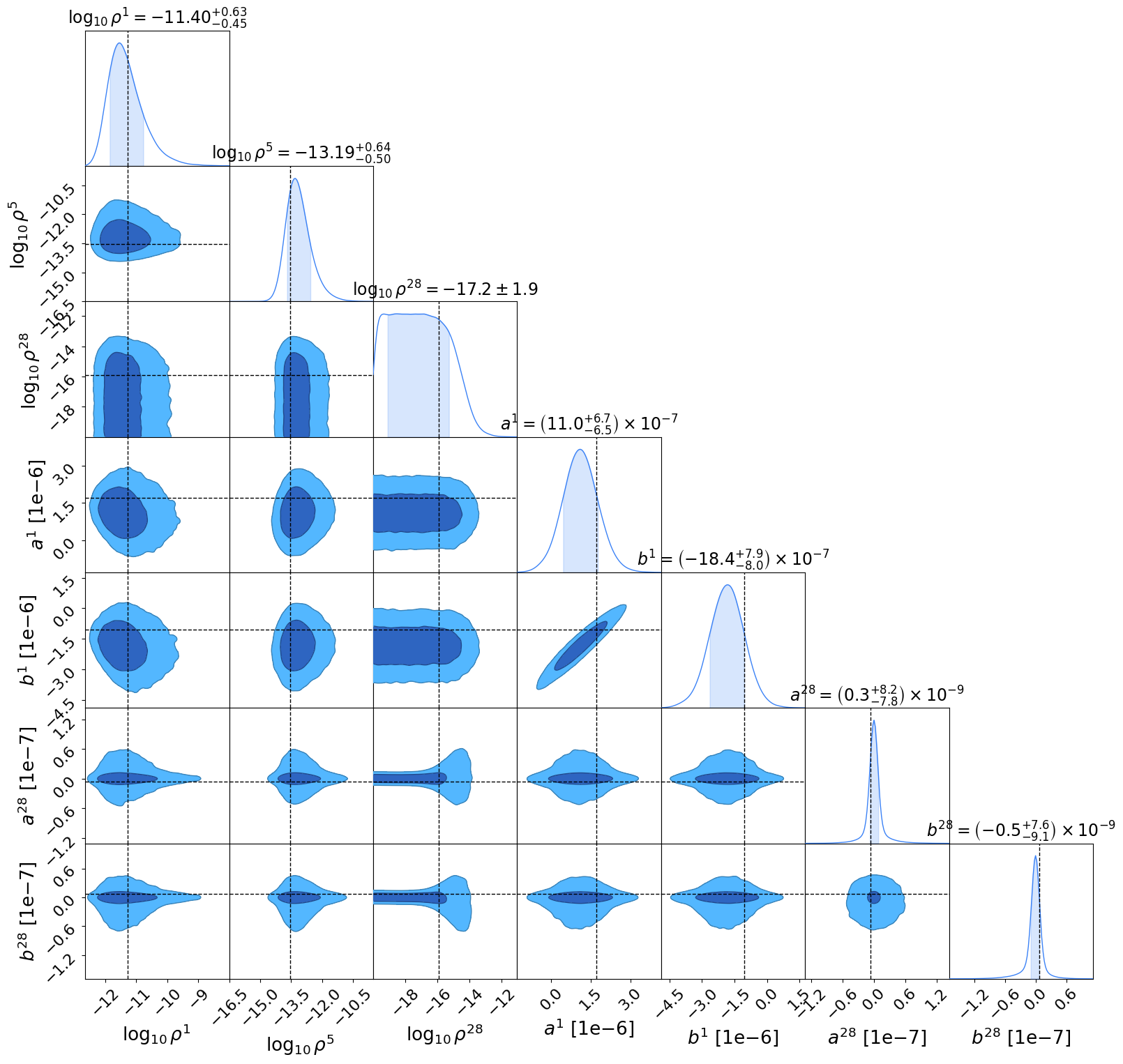}
    \caption{Samples of selected parameters obtained via HMC in a free spectral red noise analysis for a single pulsar. The first three columns correspond to the power modeled freely in various frequency bins. The remaining columns are samples over the Fourier coefficients. The ``$a$" Fourier coefficients scale sine functions and ``$b$" scale cosine. Parameter superscripts index frequency bins. Rather than an exponential funnel, a ``test tube" geometry is observed in the posterior samples plotted between hyper- and latent parameters in high frequency bins. The dashed lines are the parameter values injected into the simulated data. The summary statistics and shading of one-dimensional marginal distributions is $1\sigma$ either side the median computed with the CDF. Contours of the two-dimensional distributions enclose the $1\sigma$ and $2\sigma$ credible regions.}
    \label{fig:NealsCorner_fs}
\end{figure*}

We conduct a single pulsar red noise analysis with simulated data as described in the last section, now using a free spectral, as opposed to a power law, hyper-model. The corner plot for a selection of parameter is shown in Figure~(\ref{fig:NealsCorner_fs}). In the high frequency bins, where the posterior is prior dominated, the freely modeled power parameterizes the variance of the corresponding Fourier coefficient samples. However, we don't see an exponential narrowing funnel. Rather, a ``test tube" geometry is observed below some threshold where the variance of the Fourier coefficients is nearly uniform. Samplers with a covariance informed jump proposal will learn this distribution more effectively than Neal's funnel as jump proposals will have to be adapted less frequently. The variance of latent parameters in Neal's funnel changes exponentially, while the test tube geometry can be learned using the coarser covariance estimation of more naive samplers.

\begin{figure}
    \centering
    \includegraphics[width=0.8\linewidth]{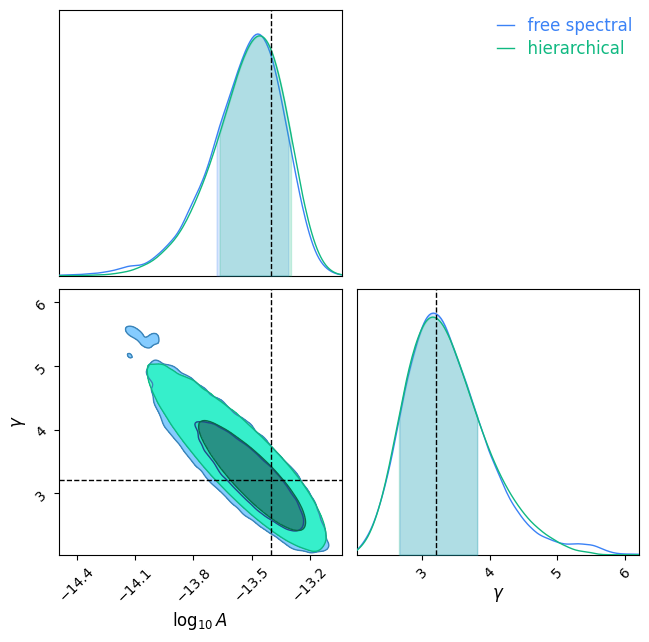}
    \caption{Samples of power law hyper-parameters from a single pulsar red noise analysis. The green posterior consists of samples from the power law hierarchical model described in Appendix~\ref{one_psr_power}. The blue posterior uses samples from the free spectral run, where the density has been estimated using normalizing flows, and reparameterized according to the power law model. The dashed lines are the parameter values injected into the simulated data. The shading of the one-dimensional marginal distributions is $1\sigma$ either side the maximum posterior. Contours of the two-dimensional distribution enclose the $1\sigma$ and $2\sigma$ credible regions.}
    \label{fig:fs_vs_power}
\end{figure}

It's possible to recover the distribution on the power law hyper-parameters (amplitude and spectral index) using the samples from a free spectral run. We can estimate the density of the free spectral samples using a kernel density estimation (KDE) or normalizing flows. Here we use \texttt{Zuko}~\cite{rozet2022zuko} which implements normalizing flows with \texttt{PyTorch}~\cite{Ansel_PyTorch_2_Faster_2024} to estimate the density of the free spectral samples. The density is then reparameterized according to the power law hyper-parameters. That is, the likelihood of a set of power law hyper-parameters is the estimated density at the corresponding point in the free spectral parameter space. The mapping between power law and free spectral power is Eq.~(\ref{eq:powerlaw}). The reparameterized (estimated) density can be sampled to obtain a distribution on the power law hyper-parameters, as shown in Figure~(\ref{fig:fs_vs_power}). This distribution is consistent with the standard power law hierarchical model described in the previous section.

\section{Comparison with \texttt{ENTERPRISE}}
We compare the results of our analysis with the standard \texttt{ENTERPRISE}~\cite{Enterprise} analysis. The analyses are conducted on the same simulated dataset, verifying we can recover consistent posterior distributions. The simulated dataset consists of 5 pulsars, each observed roughly every month for 15 years. A correlated stochastic GWB is injected in 5 frequency bins per pulsar according to the power law parameters, $\log_{10}A_B=-14.5$ and $\gamma=13/3$. We additionally inject and model a single CW signal. We will not inject nor model intrinsic pulsar RN because it is nearly identical to the stochastic GWB in form and analysis, the only difference being the inter-pulsar correlation included in the background. Moreover, the background dominates the computational cost of the analysis because the full (cross-pulsar) covariance matrix for the Fourier coefficients is not diagonal, while the intrinsic pulsar RN has a diagonal covariance matrix Eq.~(\ref{cov}) which can be factorized per pulsar. 

Our analysis differs from the standard \texttt{ENTERPRISE} analysis in two ways. First, the posterior density constructed by \texttt{ENTERPRISE} analytically marginalizes over the Fourier coefficients. Our analysis keeps the Fourier coefficients as model parameters and directly samples them via MCMC. Retaining the coefficients as model parameters makes the posterior evaluation computationally efficient, but requires sampling a high-dimensional parameter space. Second, the \texttt{ENTERPRISE} analysis explicitly evaluates the CW model over the observed TOAs. Our analysis uses the compressed Fourier representation to interpolate the CW model from sparsely sampled times to the observed TOAs. See Appendix~\ref{app:F_CW} for comparisons between our CW waveform generation and those of standard analyses.

Other than computational efficiency, the differences between our approach and \texttt{ENTERPRISE} will not effect the parameter estimation. The choice to analytically marginalize the Fourier coefficients or sample over them amounts to analytic or numeric marginalization, under which the posterior is invariant. Moreover, our CW model differs from standard approaches only in waveform generation; the likelihood implementation and sampled parameters are the same. So long as our CW waveform generation is accurate, the posterior will be identical to previous methods.

\begin{figure*}
    \centering
    \includegraphics[width=0.8\linewidth]{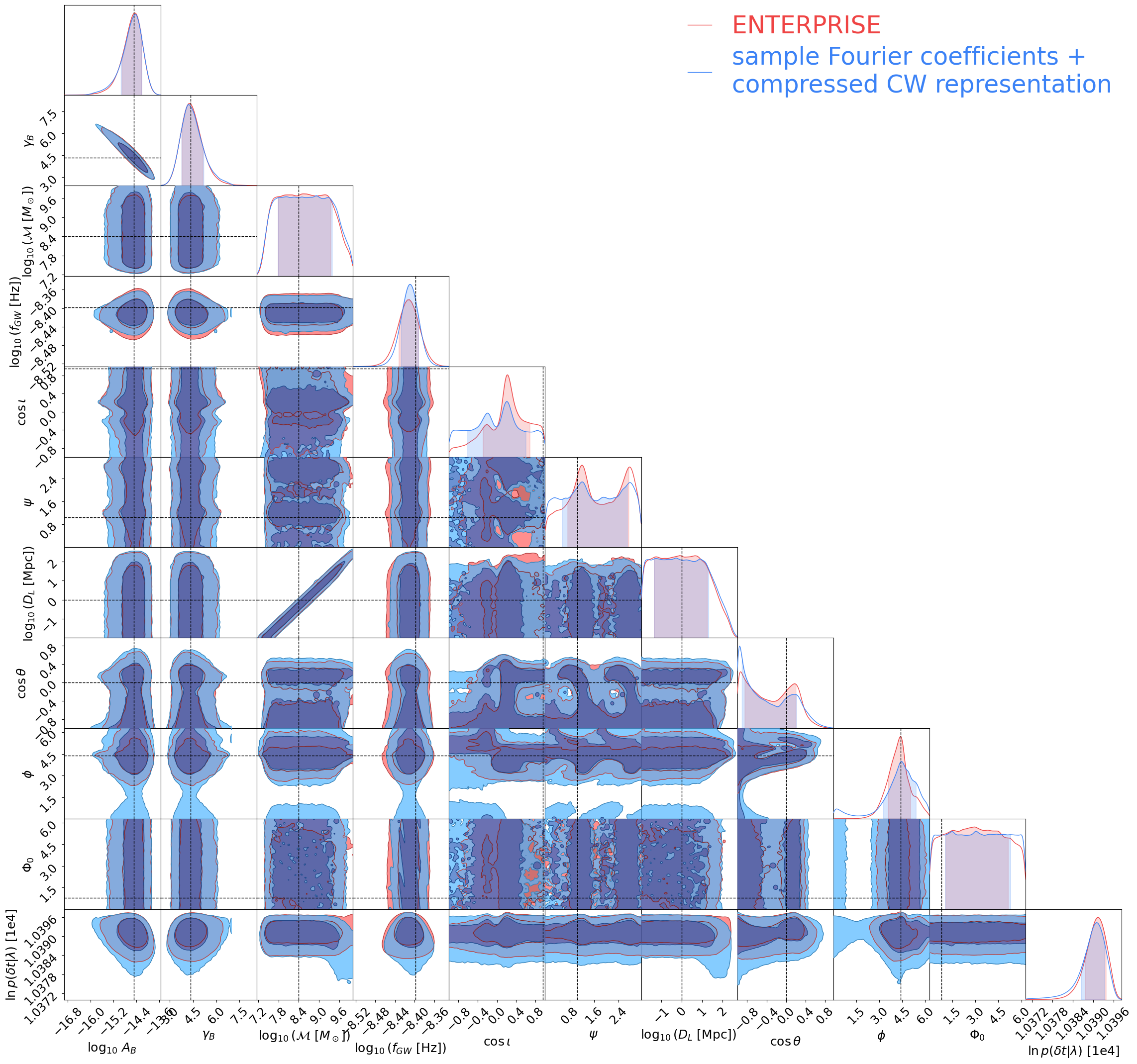}
    \caption{Samples of stochastic GWB hyper-parameters, selected CW parameters, and log-likelihood evaluations. The red posterior is constructed using the \texttt{ENTERPRISE} framework and sampled with the NANOGrav PTMCMC sampler. The blue posterior is constructed using the methods presented in this paper and sampled with HMC. The \texttt{ENTERPRISE} method analytically marginalizes over the Fourier coefficients and evaluates the CW model over the observed TOAs. The methods in this paper numerically marginalize the Fourier coefficients via direct sampling and the CW waveform is generated using the compressed Fourier representation. The dashed lines denote the injected parameter values. The shading of one-dimensional marginal distributions is $1\sigma$ either side the median computed with the CDF. Contours of the two-dimensional distributions enclose the $1\sigma$ and $2\sigma$ credible regions.}
    \label{fig:GWB_CW_Enterprise}
\end{figure*}

The posterior in the \texttt{ENTERPRISE} framework is sampled with NANOGrav's \texttt{PTMCMCSampler}~\cite{NG_PTMCMC}. The posterior constructed from the methods presented in this paper is implemented in \texttt{JAX}~\cite{JAX}, sampled using HMC from the \texttt{NumPyro} package~\cite{NumPyro1, NumPyro2}, and utilizes automatic differentiation. The posterior recovery over the stochastic GWB power law parameters and a subset of CW model parameters for these two methods is shown in Figure~(\ref{fig:GWB_CW_Enterprise}). The two posteriors over the stochastic parameters are consistent; numerical sampling of the Fourier coefficients is equivalent to analytic marginalization. The posteriors over the CW parameters do not quite match. This deviation is due to the sampler for the \texttt{ENTERPRISE} method failing to explore a secondary mode in the posterior. This secondary mode is in the sky location parameter $\phi$, but is correlated with other parameters. It is possible to resolve this secondary mode using the sampler for the \texttt{ENTERPRISE} method. Running the MCMC chain for longer, implementing custom jump proposals, or sampling hotter temperature chains would better explore the posterior. However, these sampling techniques are computationally expensive and will result in longer convergence times. Here we demonstrate the ``out-of-the-box" sampler fails to explore the posterior as effectively as HMC. We confirm this secondary mode is accessible by plotting the sampled log-likelihood values, $\ln p(\boldsymbol{\delta t}|\boldsymbol{\lambda})$, as computed by \texttt{ENTERPRISE} in Figure~(\ref{fig:GWB_CW_Enterprise}). The log-likelihood values at the secondary mode are comparable to those of the primary mode, but the \texttt{ENTERPRISE} chain fails to explore past the log-likelihood dip bridging the two modes. Besides the unexplored mode, the two posterior are in good agreement considering finite sampling.

The \texttt{ENTERPRISE} posterior is $2 + 8 + 2N_p =$ 20-dimensional because the Fourier coefficients are analytically marginalized out of the model. Our method numerically samples over the Fourier coefficients, and must sample a posterior density of dimension $2 + 2N_p N_f + 8 + 2N_p = 70$ for this example. Despite the high-dimensional parameter space, our analysis converges much faster than standard approaches due to the hyper-efficient likelihood evaluation and effective sampling with HMC. We calculate the number of effective (uncorrelated) samples per second per parameter in both approaches, the minimum of which determines the overall effective sampling rate. Our approach achieves approximately 0.065 effective samples per second, while \texttt{ENTERPRISE} achieves approximately 0.0061 effective samples per second making our method over an order of magnitude more efficient. The quoted effective samples per second are highly machine- and implementation-dependent. However, we are already able to achieve a significant speed-up on an unrealistically small dataset. As more pulsars and frequency bins are added to the model, our method will scale more efficiently than standard approaches. To demonstrate this, we consider another simulated dataset consisting \textit{only} of a stochastic GWB in 20 pulsars, each with power in 10 frequency bins. For this case our method draws independent samples at nearly 35 times the rate of the standard \texttt{ENTERPRISE} approach. For a discussion on the scaling of operation counts between methods, see the end of Section~\ref{intro}.

\bibliography{refs}

\end{document}